\documentclass{aa} 
\usepackage{graphicx}
\usepackage{natbib} 
\usepackage{txfonts}
\usepackage[dvipsnames]{xcolor}
\usepackage{soul}

\begin{document} 

\title{Tidal deformability and other global parameters of compact stars 
with strong phase transitions}
\subtitle{}

\author{M. Sieniawska\inst{1}\fnmsep
       \thanks{Corresponding author: msieniawska@camk.edu.pl}
       \and 
       W. Turcza{\'n}ski
       \inst{2}
       \and
       M. Bejger\inst{3,1} 
       \and
       J. L. Zdunik\inst{1}
       }

\institute{
Nicolaus Copernicus Astronomical Center, PAS, ul. Bartycka 18, 00-716, Warsaw, Poland
\and 
Astronomical Observatory, Adam Mickiewicz University, Pozna{\'n}, Poland
\and 
APC, AstroParticule et Cosmologie, Université Paris Diderot, CNRS/IN2P3, CEA/Irfu, Observatoire de Paris, Sorbonne Paris Cité, F-75205 Paris Cedex 13, France
}

\date{Received ...; accepted ...}

\titlerunning{Global parameters of compact stars with phase transitions}
\authorrunning{Sieniawska et al.}

 
\abstract
{Using parametric equations of state (relativistic polytropes and a simple quark bag model) 
to model dense-matter phase transitions, we study global, measurable astrophysical 
parameters of compact stars such as their allowed radii and tidal deformabilities. 
We also investigate the influence of stiffness of matter before the onset of the 
phase transitions on the parameters of the possible exotic dense phase.}
{The aim of our study is to compare the parameter space of the dense 
matter equation of state permitting phase transitions to a sub-space compatible 
with current observational constraints such as the maximum observable mass, 
tidal deformabilities of neutron star mergers, radii of configurations before 
the onset of the phase transition, and to give predictions for future observations.}
{We studied solutions of the Tolman-Oppenheimer-Volkoff equations for a flexible
set of parametric equations of state, constructed using a realistic description
of neutron-star crust (up to the nuclear saturation density), and relativistic
polytropes connected by a density-jump phase transition to a simple bag model
description of deconfined quark matter.}  
{In order to be consistent with recent observations of massive neutron 
stars, a compact star with a strong high-mass phase transition cannot have a radius 
smaller than 12 km in the range of masses $1.2 - 1.6\,M_\odot$. We also compare 
tidal deformabilities of stars with weak and strong phase transitions with 
the results of the GW170817 neutron star merger. Specifically, we study 
characteristic phase transition features in the $\Lambda_1-\Lambda_2$ relation, 
and estimate the deviations of our results from the approximate formul{\ae} for 
$\tilde{\Lambda}-R(M_1)$ and $\Lambda$-compactness proposed in the literature.  
 We find constraints on the hybrid equations of state to produce stable neutron stars on the twin branch. For the exemplary equations of state most of the high-mass twins occur for the minimum values of the density jump $\lambda = 1.33-1.54$; corresponding values of the square of the speed of sound are $\alpha = 0.7-0.37$. We compare results with observations of gravitational waves and with the theoretical causal limit and find  that the minimum radius of a twin branch is between 9.5 and 10.5 km, and depends on the phase transition baryon density. For these solutions the phase transition occurs below 0.56 [fm$^{-3}$]. }
{}
\maketitle

\section{Introduction}
\label{sect:intro}

The interior composition of  neutron stars (NSs) is still not fully known today. Neutron stars
are so compact that the density in their central regions exceeds by far the
density of the atomic nuclei. At such extreme conditions matter may exist in a form
impossible to obtain and study in  terrestrial laboratories.  One of the
possibilities of  very dense matter is the {deconfinement} of  quarks. 

Astrophysically, we study the dense matter in the interiors of NSs by measuring
their parameters (masses, radii, etc.) and compare them to the theoretical
models of a structure and equation of state (EOS). The multiple EOS models, which include a deconfined quark-matter segment, were recently proposed in \citet{Ranea2016}, \citet{Alford2017}, \citet{Kaltenborn2017}, \citet{Mellinger2017}, \citet{Christian2018} and \citet{Typel2018} \citep[see e.g.][for a review]{Buballa2014, Alford2016}. A defining characteristic of these EOS  is  the phase transition between  normal matter and  quark matter. To be consistent with the observational constraints, all the EOS have to satisfy the maximum mass $M_{max}$ criterion, which is now established to be approximately $2\,M_\odot$ (see \citealt{Antoniadis2013} for PSR J0348+0432 and \citealt{Demorest2010} for PSR J1614-2230, as well as \citealt{Fonseca2016} for the re-evaluation of the mass of the latter).

Creation of a new quark phase---the appearance of a new phase core in
the centre of a NS---may lead to a destabilisation 
of a part of the sequence of
NS configurations (where the sequence is labelled by a central EOS parameter,
such as central pressure, $P_c$). If the stability is regained further up the
sequence, it is considered  a detached (twin) branch of the compact stars (see right panel of Fig.~\ref{fig:schem} for a schematic depiction of this
situation in the $M(R)$ plane). In principle, both mass maxima (twin solutions) may exist around the maximum mass $M_{max}{\approx}2\,M_\odot$, and  is called the {high-mass twins solutions}. 

In this article we study the properties related to the twin branch of the NS using 
a simplified parametric EOS that allows us to analyse the space of NS solutions. 
Specifically, we study the influence of a strong (destabilising) phase transition 
on the tidal deformability of the NS, and apply these results to the recent measurements 
obtained by the LIGO-Virgo Collaborations from the GW170817 event \citep{Abbott2017a, Abbott2018a, Abbott2018b}. For the recent studies concerning the tidal deformability see 
\citet{Annala2017}, \citet{Most2018}, \citet{NandiC2018}, \citet{Paschalidis2018} and \citet{Raithel2018}.
We also aim to constrain how the part of the EOS {before} the phase
transition influences the parameters of the EOS {after} the phase
transition. In other words, we aim to put constraints on the observable
parameters of the NS (such as the radius that is potentially a measurable
parameter) for the masses below that at which the phase transition occurs, and
investigate how the shape of the $M(R)$ is related to the properties of the
twin branch. 

The article is composed as follows. In Sect.~\ref{sect:tov-eos} we describe 
the parametric EOS and the methods of obtaining the solutions of the hydrostatic 
equilibrium equations for the NSs. In Sect.~\ref{sect:twin-propert} we present 
the properties of NS sequences compatible with existing observational constraints, 
such as the $M_{max}$ requirement and the tidal constraints of GW170817. 
Section~\ref{sect:concl} contains discussion and summary.   

\section{Equations of state and methods}
\label{sect:tov-eos}

\begin{figure}
   \centering
   \includegraphics[width=0.95\hsize]{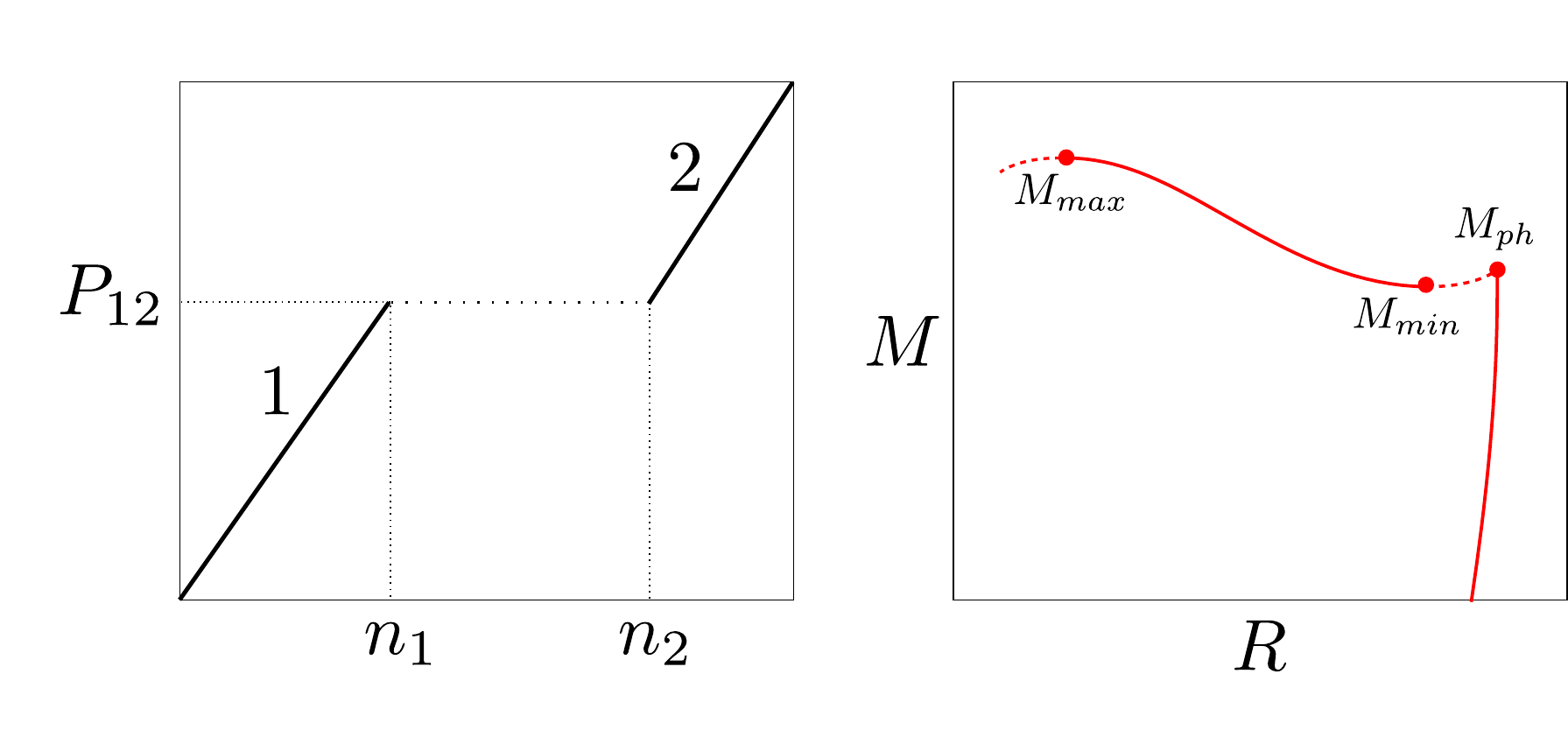}
      \caption{Left panel: Schematic EOS with a density jump phase transition
on the pressure $P$--baryon density $n_b$ plane. Right panel: Gravitational mass
$M$--radius $R$ sequence of solutions in the case of a strong (destabilising) phase
transition. We show the characteristic masses in this situation: the mass at the
phase transition density $M_{ph}$, the maximum mass $M_{max}$, and the minimum mass
$M_{min}$ at the end of the instability (decreasing part of $M(R)$ between
$M_{min}$ and $M_{ph}$ indicated with a dotted line). Stable configurations between 
the $M_{max}$ and $M_{min}$ are sometimes called the twin branch.}
         \label{fig:schem}
   \end{figure}

In order to survey the space of solutions corresponding to $M(R)$ sequences with 
high-density phase transitions, we employ a conservative approach and use the following simplified, parametric
EOS. We assume the knowledge of the low-density part of the EOS and adopt the SLy4 EOS 
description of \citet{HaenselP1994} and \citet{DouchinH2001} up to a
baryon density $n_0$, comparable to and typically larger than the nuclear
saturation density ($n_s\equiv 0.16\,\text{fm}^{-3}$). At $n_0$ a relativistic
polytrope \citep{Tooper1965} replaces the tabulated SLy4 EOS. The definition of the
pressure $P$ and the energy-density $\rho c^2$ are standard,  
\begin{eqnarray} 
P=\kappa n^{\gamma},\quad \rho c^2 = \frac{P}{\gamma -1} + nm_{b}c^2, 
\label{eq:polytrope} 
\end{eqnarray}where $\kappa$ is the pressure coefficient, $\gamma$ is the index of the
polytrope, and $m_{b}$ is the mass of the baryon
in this phase. The index $\gamma$ is a parameter of choice; consequently, by
demanding the chemical and mechanical equilibrium at $n_0$, $\kappa$ and
$m_{b}$ are fixed. The polytrope ends at a density $n_1 > n_0$, and is
connected to a simple bag EOS \citep{Chodos1974,Farhi1984}, characterising 
the quark matter. We use a linear pressure-density relation of \citet{Zdunik2000}, 
\begin{equation} 
P(\rho)=\alpha(\rho - \rho_*)c^2, \quad n(P) = n_* \left(1+\frac{\alpha+1}{\alpha}\frac{P}{\rho_*c^2}\right)^{1/(\alpha+1)}, 
\label{eq:bag}
\end{equation}  
with $\alpha$ denoting square of the speed of sound in a quark matter, and $\rho_*$ 
and $n_*$
the
energy density and baryon density of this matter at zero pressure, respectively. 
We assume that at the polytrope/bag boundary matter is softened by the
first-order phase transition defined by a density jump 
$\lambda=n_2/n_1$. Maxwell construction at this point results in a corresponding
mass-energy density jump $\rho_2/\rho_1=\lambda+(\lambda-1) P_1/\rho_1 c^2$. The values of $n_2, \rho_2$ determined by the definition of the transition point $(n_1,\rho_1,P_1)$, together with a given $\lambda$ yield, from Eq.\ref{eq:bag}, the values of $\rho_*=\rho_2-P_1/\alpha c^2$ and $n_*$. 
The schematic pressure--density relation for such an EOS is
presented in the left panel of Fig.~\ref{fig:schem}. The initial parameter
ranges are shown in Table ~\ref{tab:params}. 
\begin{table}
  \centering 
  \begin{tabular}{| c | c | c | c | c | c |} 
    \hline 
      & $n_0\ [\text{fm}^{-3}]$ & $\gamma$ 
      & $n_1\ [\text{fm}^{-3}]$ & $\lambda$ & $\alpha$ \\ \hline\hline 
    min & 0.03 & 2.25 & 0.12 & 1.3 & 0.2 \\ \hline 
    max & 0.30 & 5.50 & 0.65 & 2.2 & 1.0 \\ \hline 
  \end{tabular}
\vskip 1em  
\caption{Ranges of the poly-quark EOS parameters used in the study: $n_0$ is the
baryon density at which the realistic SLy4 ends and a relativistic polytrope
(Eq.~\ref{eq:polytrope}) with index $\gamma$ is attached. The polytrope runs
until $n_1$, at which a density jump $\lambda=n_2/n_1$ is introduced (see
Fig.~\ref{fig:schem}). The linear approximation to the quark EOS (Eq.~\ref{eq:bag}) 
characterised by the $\alpha$ parameter begins at $n_2 = \lambda n_1$.}  
\label{tab:params}
\end{table} 

Given the EOS, we solve the equations of hydrostatic equilibrium for a spherically 
symmetric distribution of mass \citep{Tolman1939,OppenheimerV1939}, 
\begin{eqnarray} 
\frac{dP(r)}{dr}&=&-\frac{G}{r^2}\left(\rho(r)+\frac{P(r)}{c^2}\right)
\left(M(r)+\frac{4\pi r^3 P(r)}{c^2}\right) \nonumber \\
&\times& \left(1-\frac{2GM(r)}{c^2r}\right)^{-1},
\label{eq:tov-dpdr}
\end{eqnarray} 
supplied with the equation for one of the metric functions, 
\begin{equation} 
\frac{d\nu(r)}{dr}= -\left(\frac{2}{P(r)+\rho(r)c^2} \right) \frac{dP(r)}{dr}, 
\label{eq:tov-dnudr}
\end{equation} 
for a spherically symmetric metric of the form
\begin{equation} 
ds^2=e^{\nu(r)} c^2 dt^2 - \frac{dr^2}{1-2GM(r)/rc^2} - r^2(d\theta^2 + \sin^2\theta d\phi^2), \label{eq:tov-metric} 
\end{equation} 
and the equation for the total gravitational mass inside the radius $r$:
\begin{equation} 
\frac{dM(r)}{dr}=4 \pi \rho(r) r^2. 
\label{eq:tov-dmdr}
\end{equation}
We solve an additional equation for the tidal deformability of the
star, defined as 
\begin{equation}
\lambda_{td} = \frac{2}{3}R^5k_2. 
\label{eq:lambda_td} 
\end{equation} 
It represents the reaction of the star on the external tidal field (such as
that in a tight binary system; e.g. \citealt{Abbott2017a}). Influence of the tidal field is obtained in the
lowest order approximation, by calculating the second (quadrupole) tidal Love
number $k_2$ \citep{Love1911}  
\begin{align} 
k_2 &= \frac{8}{5}x^5 (1-2x)^2 \bigl( 2-y + 2x(y-1)\bigr)   
\Bigl( 2x \bigl( 6 -3y +3x(5y-8)\bigr) \Bigr. \nonumber \\  
&+ 4x^3 \left( 13 -11y + x(3y -2) + 2x^2(1+y) \right) \nonumber \\  
&\left.+ 3(1-2x)^2 \bigl( 2 - y + 2x(y-1)\bigr) \ln(1-2x) \right)^{-1},  
\end{align} 
with the star's compactness $x=GM/Rc^2$, and $y$ the solution of 
\begin{align}
\frac{dy}{dr} &= -\frac{y^2}{r}-\frac{1+4\pi G r^2/c^2(P/c^2-\rho)}{(r-2GM(r)/c^2)}y  \nonumber \\
&+ \left(\frac{2G/c^2(M(r) + 4\pi r^3 P/c^2)}{\sqrt{r}(r-2GM(r)/c^2)}\right)^2 + \frac{6}{r-2GM(r)/c^2} \nonumber \\ 
&- \frac{4\pi G r^2/c^2}{r-2GM(r)/c^2}\left(5\rho+ 9P/ c^2
+\frac{\left( \rho   + P/c^2\right)^2 c^2}{ \rho dP/d\rho}\right),
\end{align}
evaluated at the stellar surface \citep{FlanaganH2008,VanOeverenF2017}. In the
following we  use the normalised value of the $\lambda_{td}$ parameter, 
\begin{equation}
  \Lambda = \lambda_{td}\left( GM/c^2\right)^{-5}. 
  \label{eq:lambda} 
\end{equation} 
Concerning the current sensitivity of the detectors \citep{Abbott2017a}, what
is actually measured is the {\it effective} tidal deformability $\tilde{\Lambda}$,
defined as 
\begin{equation} 
  \tilde{\Lambda} = \frac{16}{13} 
  \frac{\left(M_1 + 12 M_2\right) M_1^4 \Lambda_1 + 
        \left(M_2 + 12 M_1\right) M_2^4 \Lambda_2} 
  {\left(M_1 + M_2\right)^5},   
  \label{eq:lambdatilde} 
\end{equation} 
with $M_1$ and $M_2$ denoting the component masses. 
The equations are solved using a Runge-Kutta fourth-order numerical scheme with a
variable integration step (see e.g. \citealt{Press1992} for details) for a
range of central parameters of the EOS (central pressure $P_c$) to obtain
global parameters of NSs: their masses and radii in the form of the mass-radius
$M(R)$ sequence. 

\section{Features of the twin branch}
\label{sect:twin-propert}

From all the possible $M(R)$ sequences resulting from the parameter set
presented in Table~\ref{tab:params}, we select those consistent with
the following criteria:  
\begin{itemize}
  \item the maximum mass $M_{max}$ is equal to or larger than the largest currently measured 
  NS mass: $M_{max} \ge 2\,M_\odot$ \citep{Demorest2010,Antoniadis2013,Fonseca2016}; 
  \item strong (destabilising) phase transition in the EOS, i.e. an instability near the phase transition point that results in the local minimum 
  in the $M(R)$ curve and the presence of the twin branch; 
  \item unless stated otherwise, the maximum mass is located on the twin branch, i.e. $M_{max} \ge M_{ph}$ (e.g. Sect.~\ref{sect:outer} for twin sequences with $M_{max} \le M_{ph}$); 
  \item the EOS is causal (the speed of sound $\sqrt{\partial P/\partial\rho}$ is smaller 
  than the speed of light in vacuum).   
\end{itemize}

\subsection{Mass-radius diagram: minimum radius for the second branch 
and the tidal deformability} 
\label{sect:mr-and-tidal}

\begin{figure}
\centering
\includegraphics[width=\hsize]{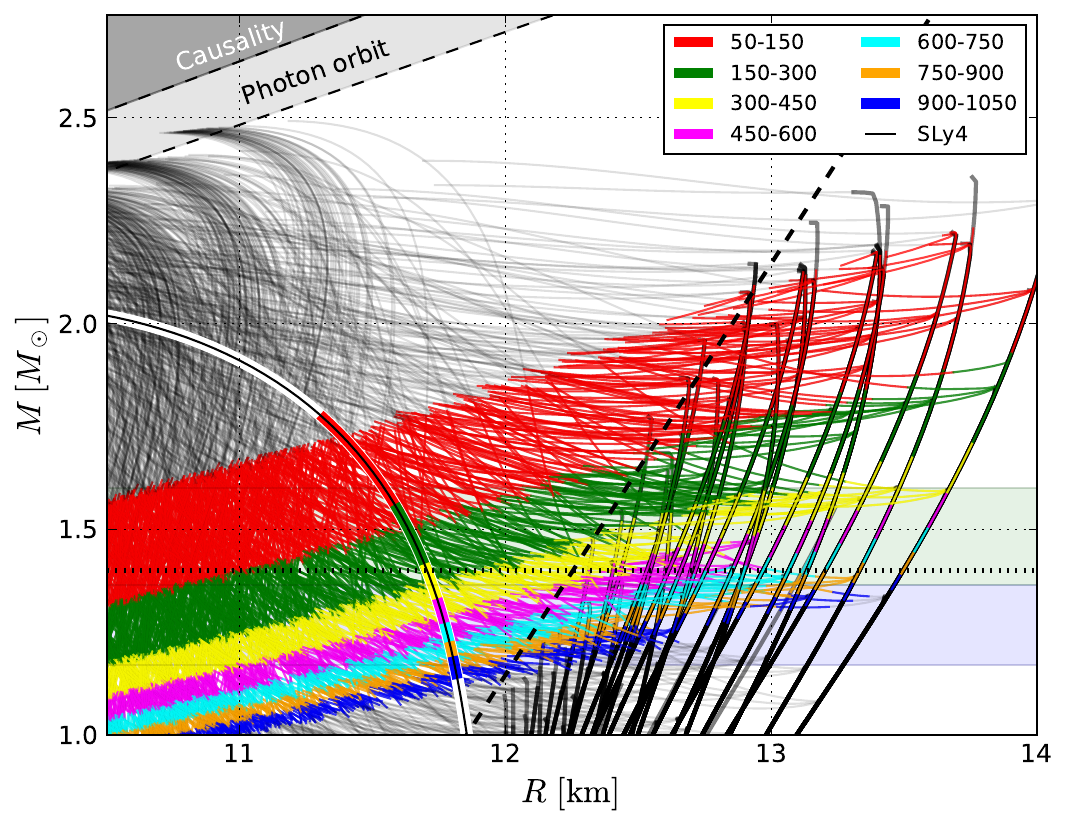}
\caption{Mass-radius diagram for the poly-quark EOS ($\alpha=1$) with the
ranges of $\Lambda(M,R)$ values  colour-coded (see inset). The
configurations  on the $M(R)$ plane may have significantly different $\Lambda$
values. This is especially visible for the black-on-white line denoting the
SLy4 EOS $M(R)$, which is placed on the top of its $\Lambda$ ranges (see
Fig.~\ref{fig:mr-lambda-ranges-zoom} for a closeup).  The dotted horizontal line
denotes $1.4\ M_\odot$. The dashed inclined black line denotes an approximate
division between the purely polytropic $M(R)$ sequences and sequences
containing quark core (see Eq.~\ref{eq:minrad}).  The pale green and blue bands
correspond to the component mass ranges estimated in the low-spin prior case of
GW170817 \citep{Abbott2017a}. For reference, in the upper left corner of the
plot we indicate the regions excluded by the requirement of the EOS causality
(speed of sound in the dense matter less than the speed of light in vacuum,
\citealt{Haensel1999}), and the photon orbit ($3GM/Rc^2$) in the Schwarzschild
case.} 
\label{fig:mr-lambda-ranges}
\end{figure}

\begin{figure}
\centering
\includegraphics[width=\hsize]{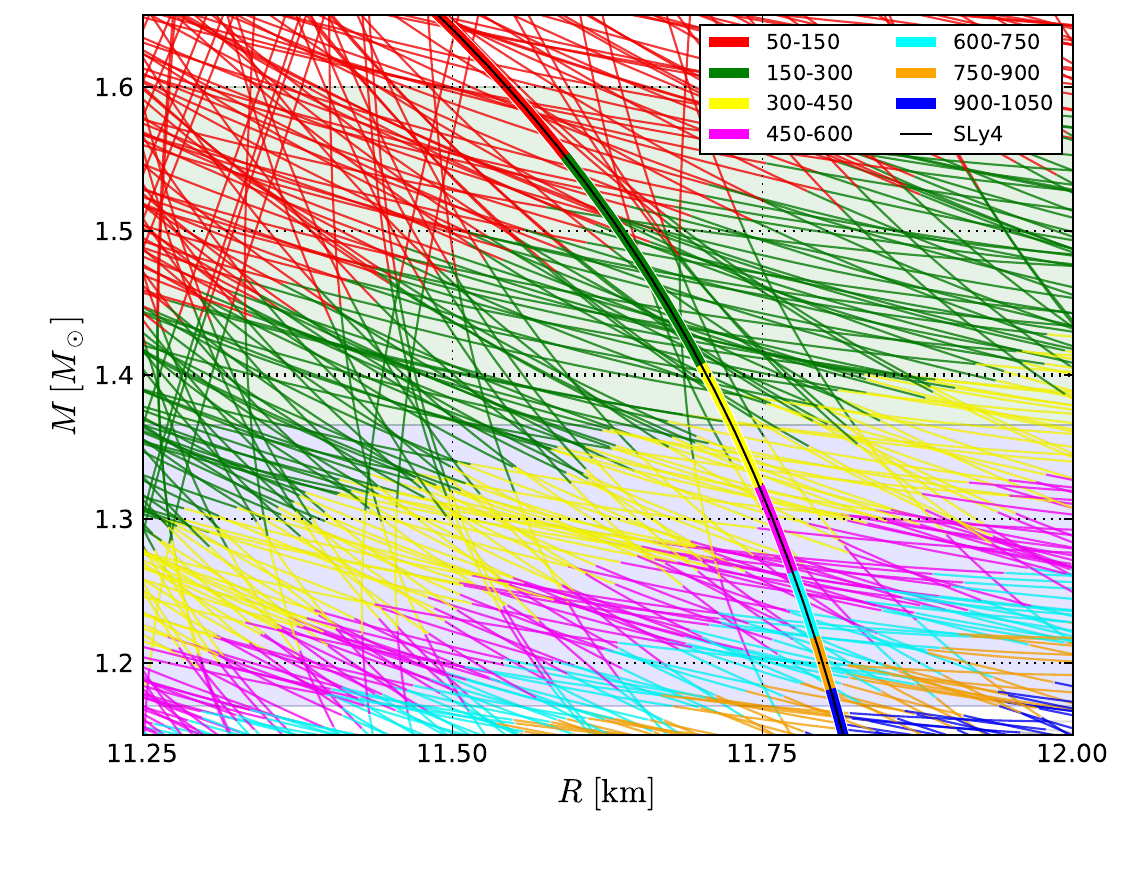}
\caption{Zoomed-in image of Fig.~\ref{fig:mr-lambda-ranges} around the SLy4 EOS $M(R)$ sequence 
showing the differences in $\Lambda$ for the configurations of the same $M$ and $R$.} 
\label{fig:mr-lambda-ranges-zoom}
\end{figure}

\begin{figure}
\centering
\includegraphics[width=\hsize]{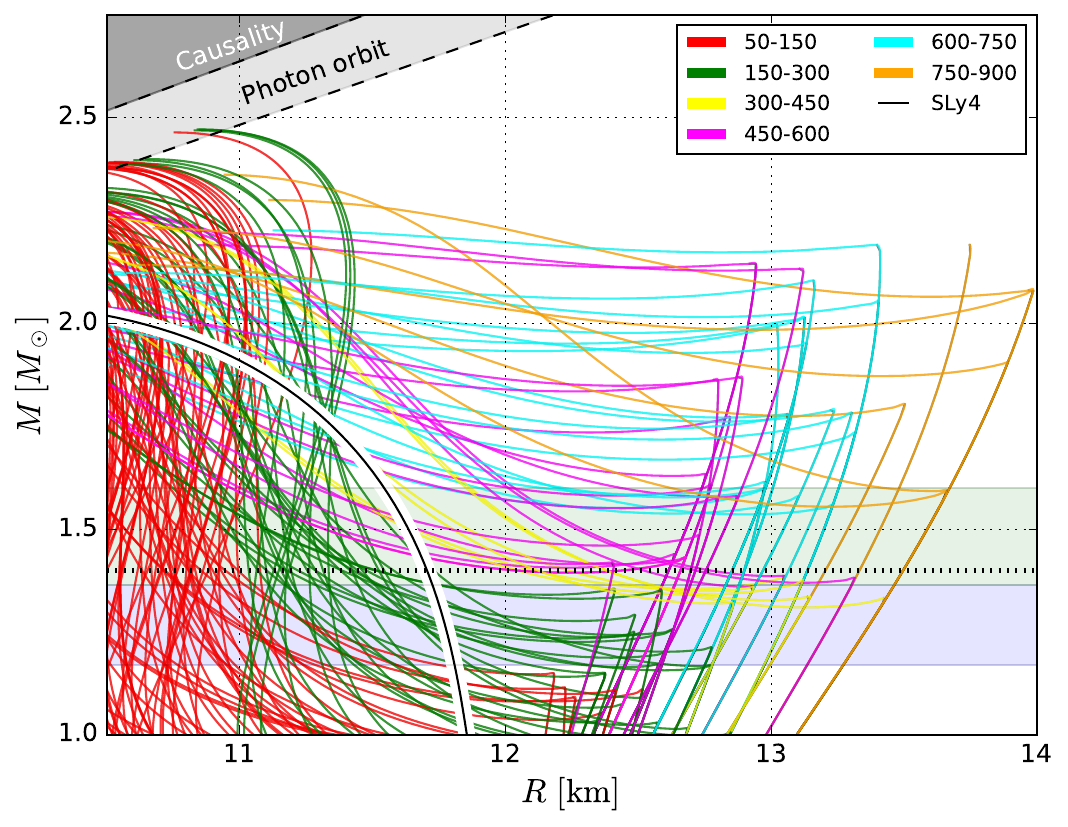}
\caption{Mass-radius diagram for the poly-quark EOS ($\alpha=1$) with the
curves colour-coded according to their $\Lambda(M_{1.4})$ values. The black-on-white
solid line denotes the SLy4 EOS $M(R)$. The regions with different
$\Lambda(M_{1.4})$ overlap each other, also because one sequence may have
several values of $\Lambda(M_{1.4})$ due to the nature of the strong phase
transition around $M=1.4\ M_\odot$.  The non-trivial behaviour of $M(R)$ hinders the
separation of the EOS on the $M(R)$ plane with respect to the values of
$\Lambda(M_{1.4})$ they yield (as proposed in \citealt{Annala2017}).}  
\label{fig:mr-lambda14-ranges}
\end{figure}

We are interested in establishing a lower limit on the radius $R$ at densities
below the phase transition point for the EOS compatible with the twin branch
scenario. Since it is likely that a large fraction of the NS population has 
masses below $M_{max}=2 M_\odot$, the future measurements of their radii
will be indicative of whether the EOS is able to support a phase transition
at higher densities (high-mass twins scenario) in a way compatible with the observations of the massive NSs.
In order to obtain the lower bound on the radius, we adopt the $\alpha$
parameter in the quark phase equal to its extreme value, $\alpha=1$.
Characteristic shapes of the $M(R)$ relations are presented in
Fig.~\ref{fig:mr-lambda-ranges}, where  selected 
regions of the $\Lambda$ parameters in the $M(R)$ plane are also shown. The values 
of $\Lambda$ decrease with $M$ and increase with $R$ thanks to the strong scaling with 
these values. A limit of 900 corresponds to the lower edge of the orange region. 
In principle, we can construct a specific EOS corresponding to limiting  
$R_{ph}$ or $M_{ph}$ still consistent with this bound. We conclude that the minimum radius R on the polytropic branch of the M(R) sequence that still allows for the twin branch scenario cannot be smaller than 12 km with the adopted SLy4 crust, for the astrophysically interesting range of masses (${\simeq}1.0 - 2.0 M_\odot$). Softer pre-phase-transition polytropes which give $R<12$~km 
result either in $M_{max}<2\,M_\odot$, or in $M_{ph} > M_{max}$.  In
addition to the polytropic index $\gamma$, factors that determine the radius
for stars below the high-density twin branch segment are the densities $n_0$ at
which the polytrope connects to the SLy4 EOS (low-density SLy4-polytrope
connection results in larger radii) and its ending point $n_1$.   This simple
exercise shows that the measurements of radii in this range of masses disclose
information pertinent to our understanding of the EOS at densities above the
nuclear saturation density. 

\begin{figure}
\centering
\includegraphics[width=\hsize]{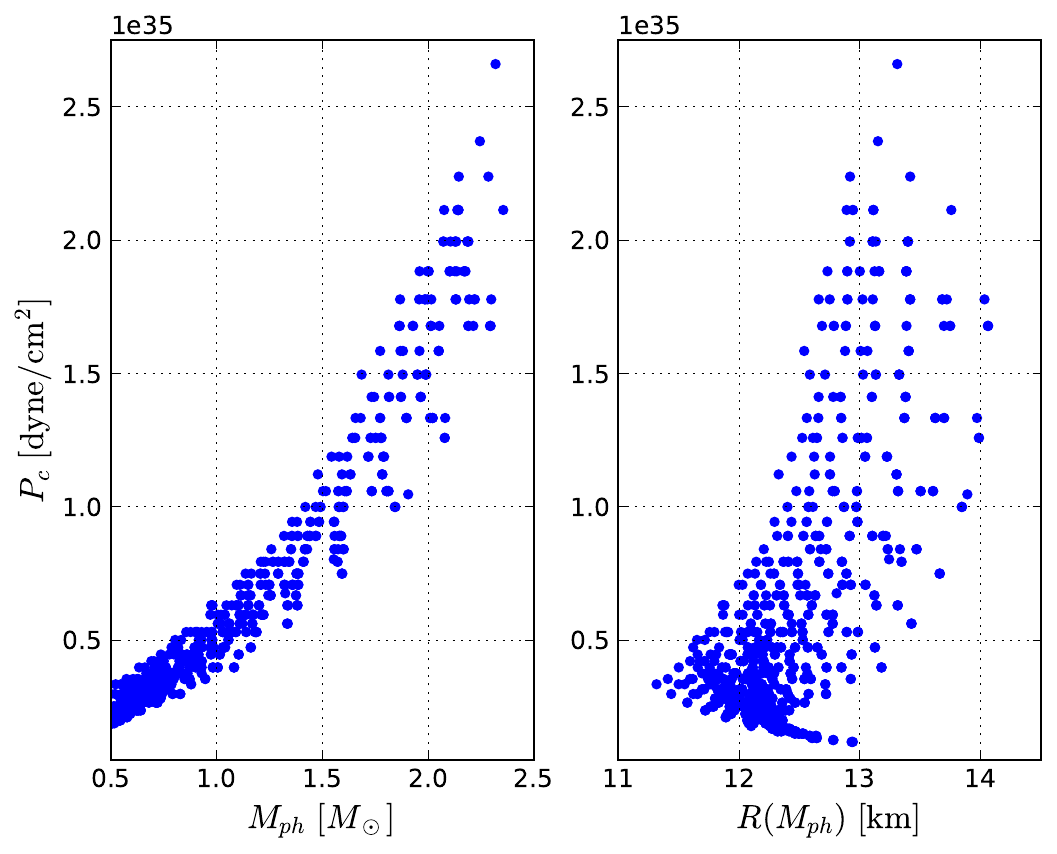}
\caption{Central pressure $P_c$ vs. mass and radius $M_{ph}$ and $R(M_{ph})$ 
at the end of the polytropic branch (corresponding to the 
phase transition point).} 
  \label{fig:mpc-rpc}
\end{figure}

\begin{figure}
\centering
\includegraphics[width=\hsize]{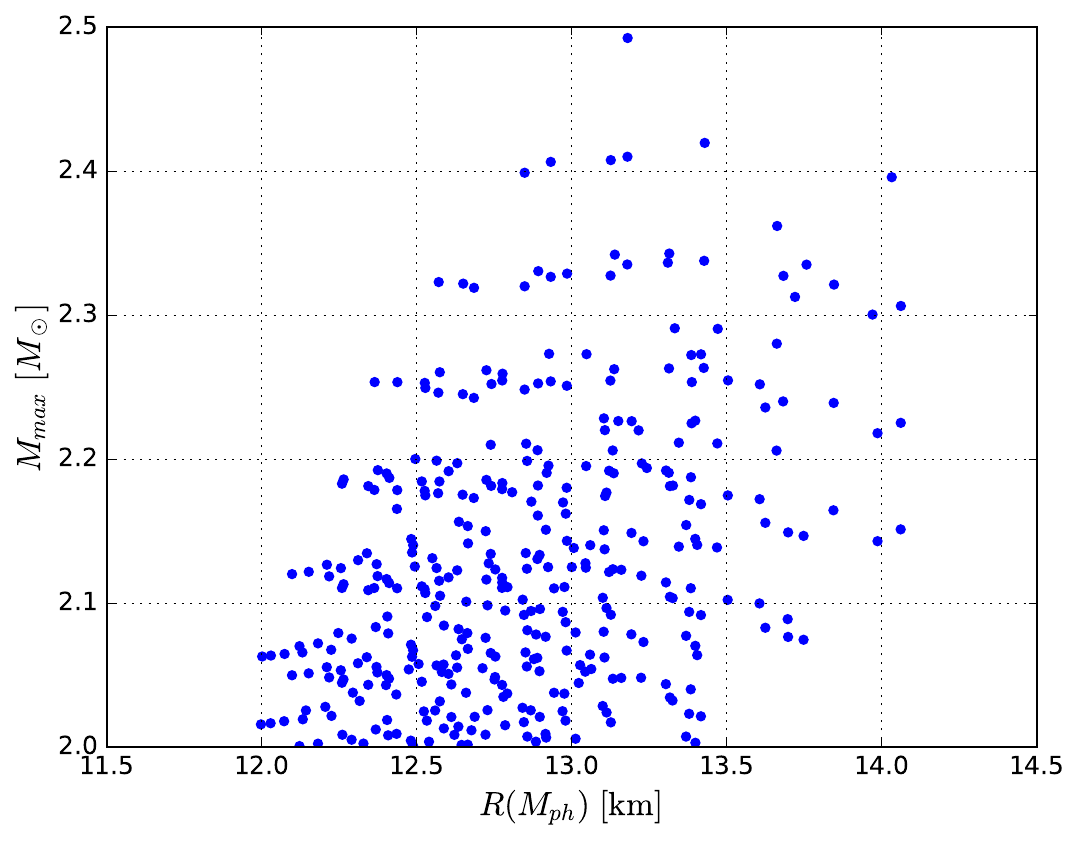}
\caption{Maximum mass on the twin branch $M_{max}$ vs. radii at the phase transition point $R(M_{ph})$.}
  \label{fig:mmax-rph}
\end{figure}

The thick dashed black line in Fig.~\ref{fig:mr-lambda-ranges} 
approximates the minimum radius of the configurations before the phase transition, 
\begin{equation} 
R = R_{SLy4}(1M_{\odot}) + (M/M_{\odot}-1)\ \mathrm{km}, 
\label{eq:minrad} 
\end{equation} 
with $R_{SLy4}(1M_{\odot})=11.85$ km. For comparison, we also plot the SLy4 EOS
curve, which resides entirely in the quark-core part of the $M(R)$ diagram.
Figure~\ref{fig:mpc-rpc} shows the relation between the central pressure
$P_c$ for configurations with $M_{ph}$ and $R(M_{ph})$ at the end of the polytropic
branch; the left-side boundaries of the scatter plots correspond to the minimum radius 
line in Fig.~\ref{fig:mr-lambda-ranges}; for example, stars at the polytropic branch 
with radii smaller than 12 km and corresponding masses smaller than $1.2\ M_\odot$ 
cannot exist with central pressures larger than approximately $7.5\times 10^{34}$ 
dyne/cm$^2$. The relation between $M_{max}$ and $R(M_{ph})$ is shown in Fig.~\ref{fig:mmax-rph}. Small radii at the transition point support lower maximum masses on the twin branch.
    

\subsection{Tidal deformability of stars with weak and strong phase transitions}
\label{sect:td}

We now study the tidal deformability $\Lambda$ parameters
(Eq.~\ref{eq:lambda}) using a subset of the parametric EOS. The ranges of 
astrophysically interesting $\Lambda$ parameters are indicated in
Fig.~\ref{fig:mr-lambda-ranges}. We note that it is virtually impossible 
to define a clear boundary between different $\Lambda$ ranges on the $M(R)$ plane. 
As the value of $\Lambda$ depends sensitively on the EOS of dense matter, it cannot be 
treated as a simple function of $M$ and $R$ only. This is
particularly visible for the comparison of the SLy4 EOS in
Fig.~\ref{fig:mr-lambda-ranges-zoom}. Here, depending on the EOS, configurations 
with the same compactness $GM/Rc^2$ but visibly different $\Lambda$ occupy the 
same position on the $M(R)$ diagram, e.g. for $M=1.32\ M_\odot$ and $R=11.75$ km, 
the end of the magenta segment yielding $\Lambda=450$ is located in the middle of 
the $\Lambda=300-450$ range for the parametric EOS. 

Figure~\ref{fig:mr-lambda14-ranges} shows a $M(R)$ diagram with the EOS
  colour-coded   according to their values of $\Lambda(M_{1.4})$. Here
the situation is even less clear in terms of defining the $\Lambda$ on the $M(R)$ 
plane, also because of the non-monotonic behaviour of $M$ due to the strong
phase transition: a given EOS may yield several values of $\Lambda(M_{1.4})$
corresponding to different $R$. We conclude that representing
$\Lambda(M_{1.4})$ on the $M(R)$ diagram (as proposed by e.g. \citealt{Annala2017}) 
is not the best way of uncovering the relation between the $\Lambda$ and other 
EOS functionals. Our results are based on a relatively simple model of the EOS; adding 
more degrees of freedom will only further complicate this picture.  

Using the selected EOS from Fig.~\ref{fig:mr-lambda-ranges} as an example, we
discuss the tidal deformability $\Lambda$ for the strong-phase transition EOS
by applying a measurement of the chirp mass, 
\begin{equation} 
  \mathcal{M}= \frac{(M_1M_2)^{3/5}}{(M_1 + M_2)^{1/5}},  
  \label{eq:chirpmass} 
\end{equation}
and the component mass ranges from the one binary NS merger observed so far in gravitational waves, 
the GW170817 event \citep{Abbott2017a}. We adopt the central value of the 
chirp mass, $\mathcal{M}=1.188\ M_\odot$, and component masses in the low-spin 
prior estimation case: $M_1 \in (1.36, 1.60)$, $M_2 \in (1.17,
1.36)$ \citep{Abbott2017a}. An improved analysis that gives consistent results is described in \citet{Abbott2018a,Abbott2018b}, where revised values for the low-spin prior of the chirp mass and component masses are $\mathcal{M}=1.186\ M_\odot$, $M_1 \in (1.36, 1.60),$ and $M_2 \in (1.16,1.36)$. A recent correction of the misprint reveal that $\tilde{\Lambda}\lesssim 900$ (Table IV in \citealt{Abbott2018b}), instead of $\tilde{\Lambda}\lesssim 800$ (\citealt{Abbott2017a}). Nevertheless, due to the very small differences between initial and re-evaluated estimations of the mentioned values, and the fact that their estimation errors overlap (in the case of tidal deformability values, errors of the estimations are considerable and depend on the waveform model, e.g. for the symmetric TaylorF2 waveform model $\tilde{\Lambda} = 340^{+580}_{-240}$ and for HPD PhenomDNRT $\tilde{\Lambda} = 300^{+430}_{-230}$, according to \citealt{Abbott2018b}), we decided to use the results from the detection paper. 

\begin{figure}
\centering
\includegraphics[width=\hsize]{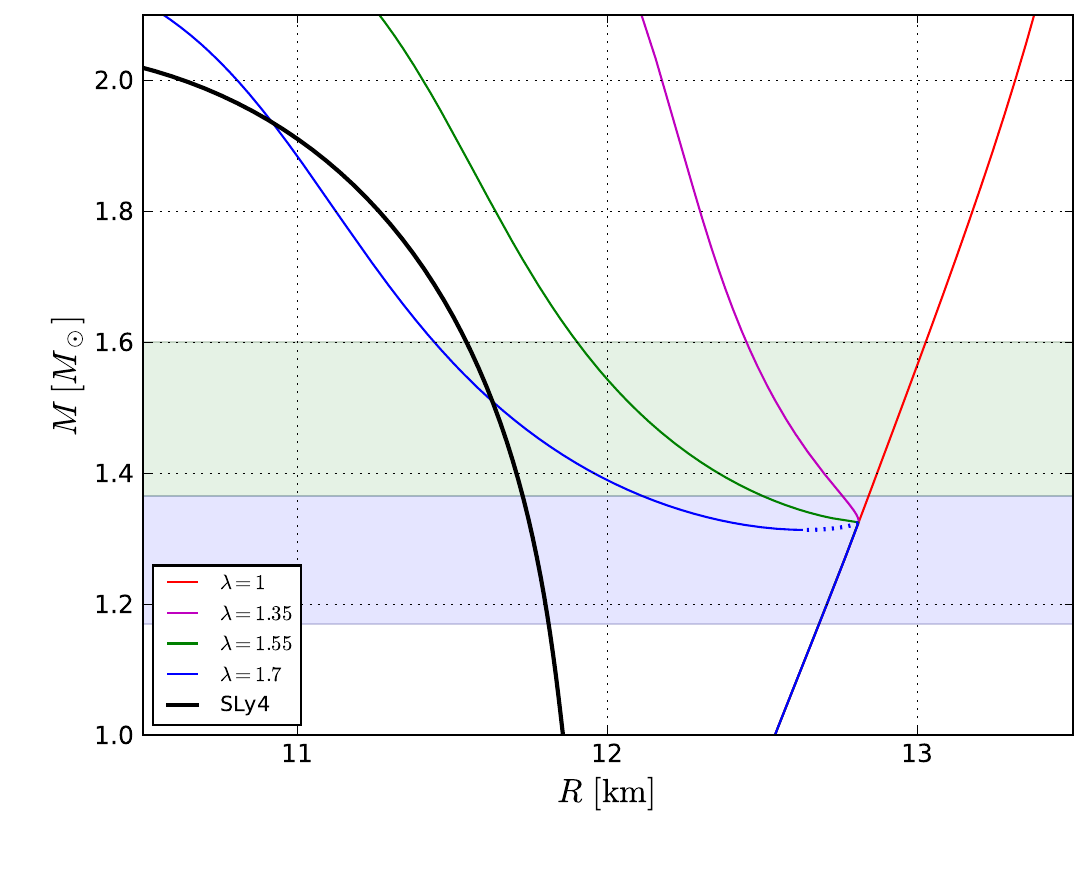}
\caption{Mass-radius relations for the polytropic-quark EOS with realistic SLy4
crust. Selected $M(R)$ relations correspond to a polytropic segment of
$\gamma=4.5$, connected to the SLy4 crust at $n_0=0.21$ fm$^{-3}$ and to a
linear EOS of $\alpha=1$ (Eq.~\ref{eq:bag}) at $n_1=0.335$ fm$^{-3}$. The
curves differ by the density jump $\lambda=n_2/n_1$, as indicated in the plot.
Regions destabilised by the phase transition are indicated by dotted segments.
The black line indicates the SLy4 EOS. The pale green and blue bands correspond to the mass
ranges estimated in the low-spin prior case of GW170817 \citep{Abbott2017a}.} 
  \label{fig:mr_phvsnoph}
\end{figure}

To visualise the influence of a phase transition softening on $\Lambda$, we
manufacture a selection of the EOS that differ in the size of the density jump
$\lambda=n_2/n_1$, and otherwise share their parameters ($M(R)$ diagrams 
shown in Fig.~\ref{fig:mr_phvsnoph}). The phase transition
from the polytropic segment of $\gamma=4.5$ to the linear EOS of $\alpha=1$ occurs
at $n_1=0.335$ fm$^{-3}$ (crust-core transition from the SLy4 EOS to a
polytropic segment occurs at $n_0=0.21$ fm$^{-3}$). The parameters are selected
such that the onset of the softening occurs within the range of masses
estimated for the GW170817 event \citep{Abbott2017a}. 

\begin{figure}
\centering
\includegraphics[width=\hsize]{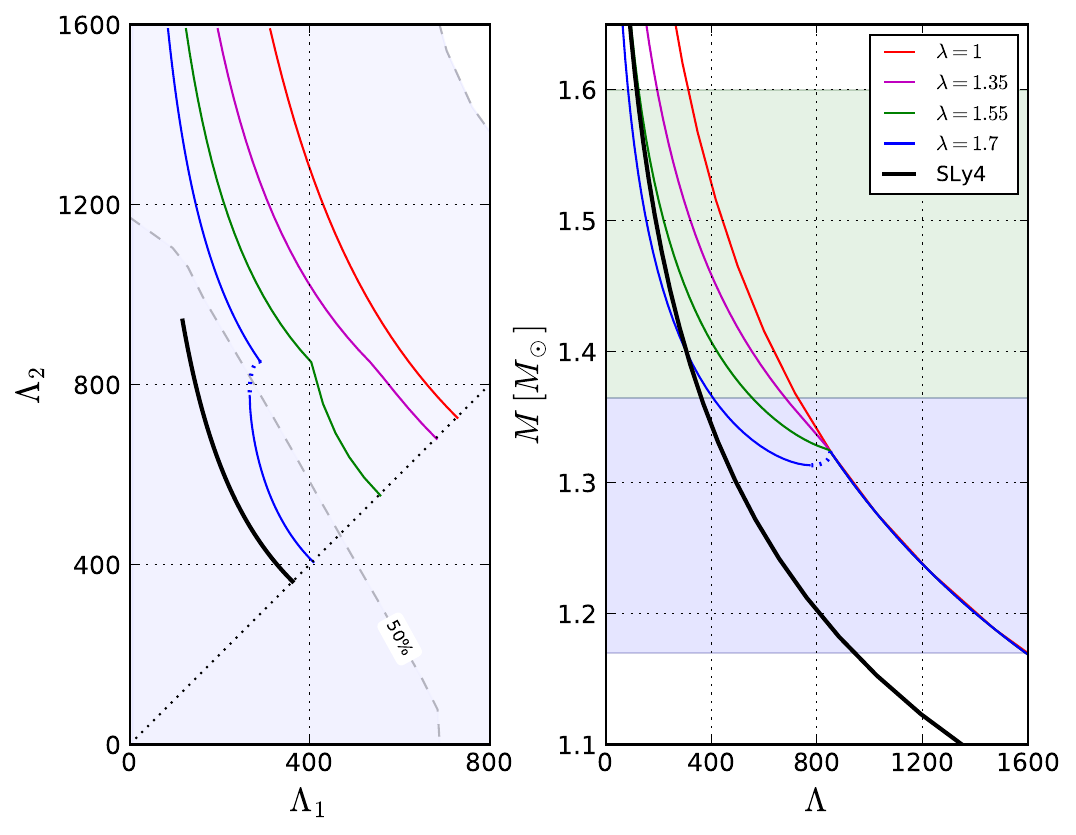}
\caption{$\Lambda_2(\Lambda_1)$ (left panel) and $M(\Lambda)$ (right panel)
relations for the $M(R)$ sequences from Fig.~\ref{fig:mr_phvsnoph}.  The values
of the $\Lambda$ parameter are based on the measurements of the chirp mass and the
low-spin prior estimates of the component masses in the binary NS merger
GW170817 \citep{Abbott2017a}. Shaded areas in the left panel denote the
estimated 50\% and 90\% confidence regions corresponding to the measurement 
\citep{Abbott2017a}.}
  \label{fig:ll_ml_phvsnoph}
\end{figure}

In Fig.~\ref{fig:ll_ml_phvsnoph} we plot the $\Lambda_1-\Lambda_2$ relations for 
the component masses of the GW170817 event by assuming the estimated central value of 
the chirp mass, $\mathcal{M}=1.188\ M_\odot$. Depending on the strength of the phase 
transition (measured by the size 
of the density jump $\lambda$), the $\Lambda_1-\Lambda_2$ relation may 
exhibit a non-monotonic behaviour, as shown in the left panel of 
Fig.~\ref{fig:ll_ml_phvsnoph} for the blue curve. This feature is a direct 
consequence of a non-monotonic behaviour of the gravitational mass $M$ as a function 
of the tidal deformability $\Lambda$ (right panel of Fig.~\ref{fig:ll_ml_phvsnoph}), 
which is connected to the instability caused by a sufficiently strong phase transition. 
Detecting this feature in the incoming observations of NS mergers will clearly signal 
a dense-matter softening in the range of central parameters corresponding to NS 
merger component masses.  

\begin{figure}
\centering
\includegraphics[width=\hsize]{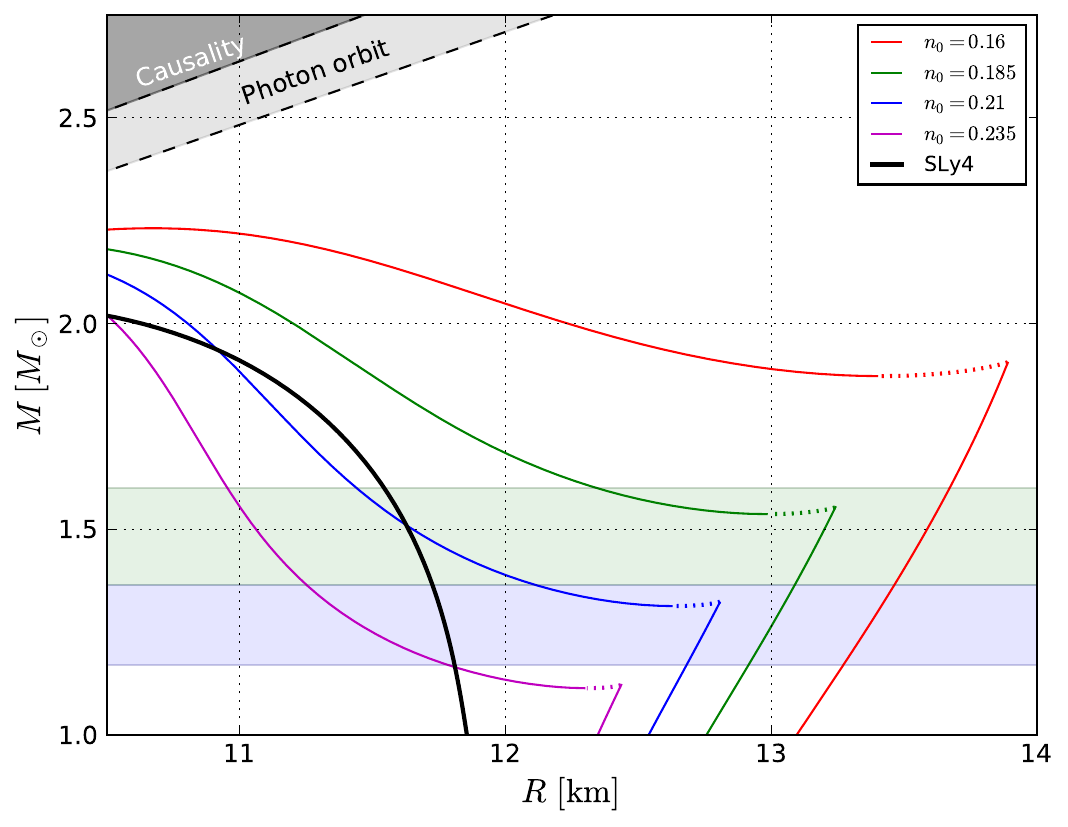}
\caption{Mass-radius relations for the polytropic-quark EOS with realistic SLy4
crust. Selected $M(R)$ relations correspond to $\gamma=4.5$, connected to the
SLy4 crust at the baryon densities $n_0$ indicated on the plot (magenta EOS at
$n_0=0.235$ fm$^{-3}$, blue EOS at $n_0=0.21$ fm$^{-3}$, green EOS at
$n_0=0.185$ fm$^{-3}$, red EOS at $n_0=0.16$ fm$^{-3}$). In all the selected
cases, the quark EOS starts at $n_1=0.335$ fm$^{-3}$, the density jump is
$\lambda=1.7$, and $\alpha=1$ (Eq.~\ref{eq:bag}). Regions destabilised by the
phase transition are indicated by dotted segments. The black line denotes the SLy4
EOS. The green and blue bands correspond to the mass ranges estimated in the
low-spin prior case of GW170817 \citep{Abbott2017a}.}  
\label{fig:mr_4quark_sly}
\end{figure}

\begin{figure}
\centering
\includegraphics[width=\hsize]{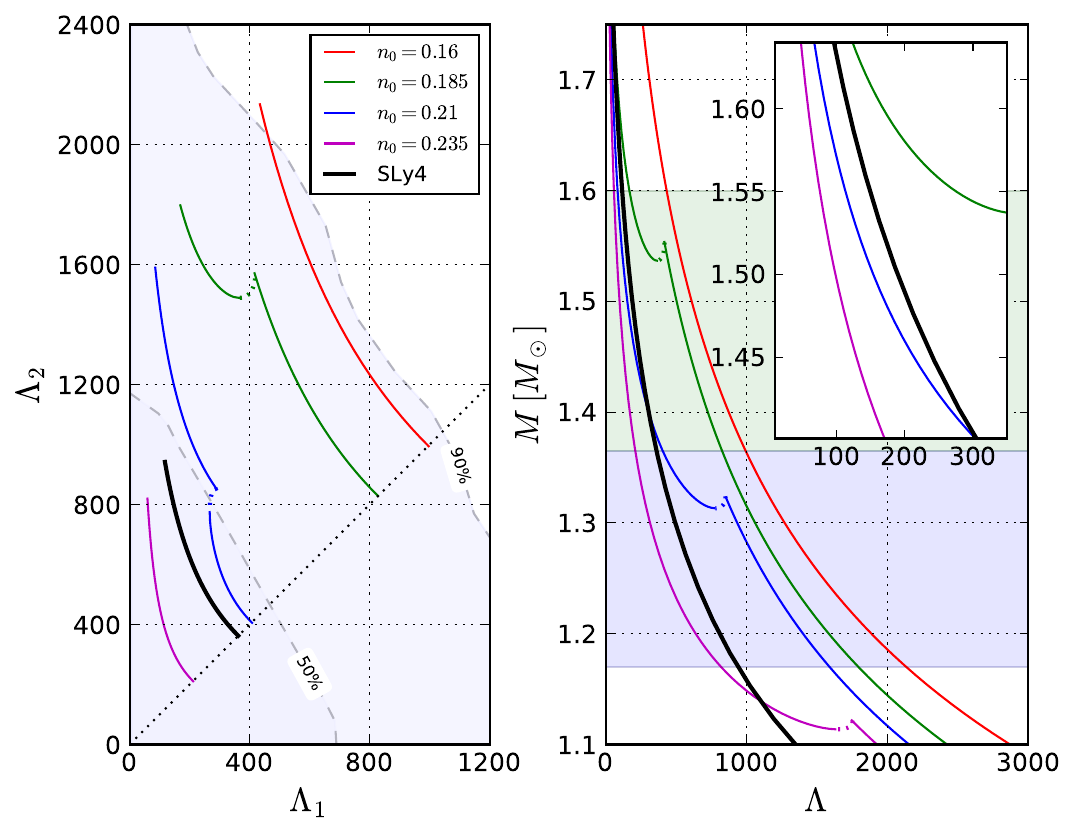}
\caption{$\Lambda_2(\Lambda_1)$ (left panel) and $M(\Lambda)$ (right panel)
relations for the selected twin branch sequences, with  colour-coding and symbols 
as in Fig.~\ref{fig:mr_4quark_sly}.} 
\label{fig:ll_ml}
\end{figure}

Additionally, we produced a set of four strong phase-transition $M(R)$ 
curves, displayed in Fig.~\ref{fig:mr_4quark_sly}: their common parameters are $\gamma=4.5$,
$n_1=0.335$ fm$^{-3}$, $\lambda=1.7$, $\alpha=1$. The EOS differ only in the
values of the SLy4-polytrope density matching point, $n_0=0.16,\ 0.185,\ 0.21,\
0.235$ fm$^{-3}$, respectively. These $n_0$ values were selected so that the phase
transition mass $M_{ph}$ is placed below, above, or within the ranges of the
component masses for the low-spin prior estimation of the GW170817 event 
\citep{Abbott2017a}. The $\Lambda_1-\Lambda_2$ relations are displayed 
in Fig.~\ref{fig:ll_ml}. Specifically, the green and blue curves are mirror 
rotated versions of themselves, reflecting the fact that the phase transition 
occurs in the range of $M_1$ or $M_2$, respectively. 

When both configurations (twin case) are present in the relevant range of masses (e.g. blue and green curves in Fig.~\ref{fig:mr_4quark_sly} and in the right panel of Fig.~\ref{fig:ll_ml}), the twin branch gives more compact $M(R)$ configurations in comparison to
those on the polytropic branch, their $\Lambda$ values are in general smaller,
and thus  preferable from the point of view of the GW170817 observations
\citep{Abbott2017a}. Likewise, larger radii before the phase transition onset
(e.g. red curve in Fig.~\ref{fig:mr_4quark_sly}) are generally less favoured  by
the observation of GW170817. This may put limits on the proposed high-mass twin
branch solutions in which the onset of a phase transition occurs near $2\
M_\odot$ \citep{Benic2015,Kaltenborn2017,Typel2018}.
Figure~\ref{fig:mr-lambda-ranges} visualises the requirement for a high-mass twin solution and the radius:  the higher the $M_{ph}$, the larger
the radius at $M_{ph}$. For $M(R)$ dependencies presented in Fig.~\ref{fig:mr-lambda-ranges} in the very large range of masses radius increases
with mass;  this property is typical for a very stiff EOS. As a consequence, high-mass twin configurations in the
range of GW170817 masses (before the phase transition) have larger radii and 
larger $\Lambda$ than lower mass twins. 

\begin{figure}
\centering
\includegraphics[width=\columnwidth]{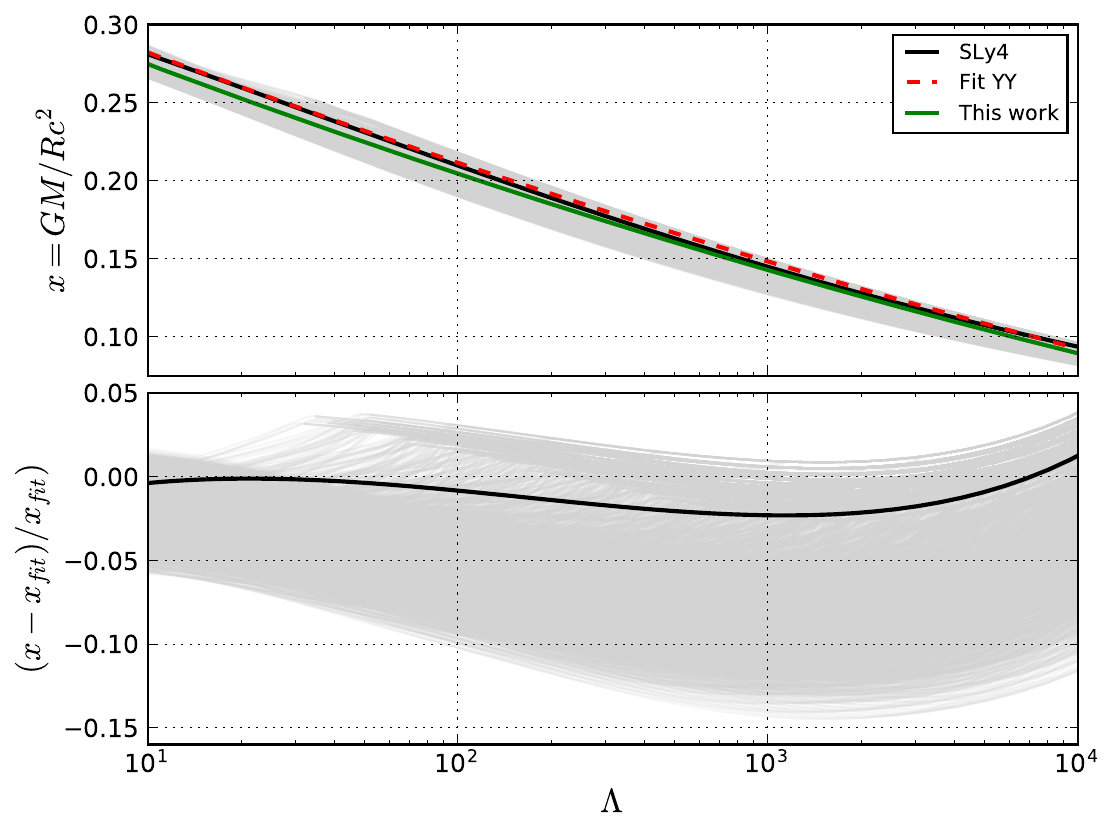}
\caption{Compactness $GM/Rc^2$ as a function of $\Lambda$ for the $\alpha=1$ EOS 
from Figs.~\ref{fig:mr-lambda-ranges}--\ref{fig:mr-lambda14-ranges} (grey lines). The black line denotes 
the SLy4 EOS result. The dashed red line is the fitting formula of \citet{YagiY2017}, 
whereas the green line is the best fit to the values described in the text. 
The lower panel shows the relative difference between the \citeauthor{YagiY2017} formula 
and the compactness values resulting from the $M(R)$ curves.} 
\label{fig:lambda_comp}
\end{figure}

The twin branch segments lie in the region compatible with the SLy4 EOS.
Even though the magenta EOS ($n_0=0.235$ fm$^{-3}$) and the blue EOS
($n_0=0.21$ fm$^{-3}$) mass-radius relations cross the SLy4 EOS $M(R)$
sequence, their $\Lambda$ values differ in these points, as shown in the right
panel of Fig.~\ref{fig:ll_ml}. For the SLy4-magenta EOS crossing at $M\approx
1.17\ M_\odot$, the $\Delta\Lambda\approx 120$ (to be compared with the value
of $\Lambda_{Sly4}\approx 970$ at this point) and $\Delta\Lambda\approx 35$ for
the SLy4-blue EOS crossing at $M\approx 1.51\ M_\odot$ ($\Lambda_{Sly4}\approx
190$ at this point). Generally, configurations with the same $M$ and $R$ (same
compactness $M/R$) have notably different tidal deformabilities depending on
their EOS. We plot the set of configurations from
Fig.~\ref{fig:mr-lambda-ranges} on the $\Lambda$-compactness $x=GM/Rc^2$ plane
(Fig.~\ref{fig:lambda_comp}), and compare the results with the fitting formula
proposed in \citet{Maselli2013} and recently updated in \citet{YagiY2017},
$x_{fit}=\sum_i a_i(\log(\Lambda))^i$ (see Eq. 78 and Fig.~15 in
\citealt{YagiY2017}). The deviation from the \citeauthor{YagiY2017} fit is, for
the EOS used here, typically larger than 5\% and increases above 10\% for
$\Lambda>100$. The best fit to our set of EOS 
has the following parameters: $a_0 = 0.353$, $a_1 =
-0.0359$, and $a_2 = 0.000790$ (indicated by  a green line in
Fig.~\ref{fig:lambda_comp}).

\begin{figure}
\centering
\includegraphics[width=\hsize]{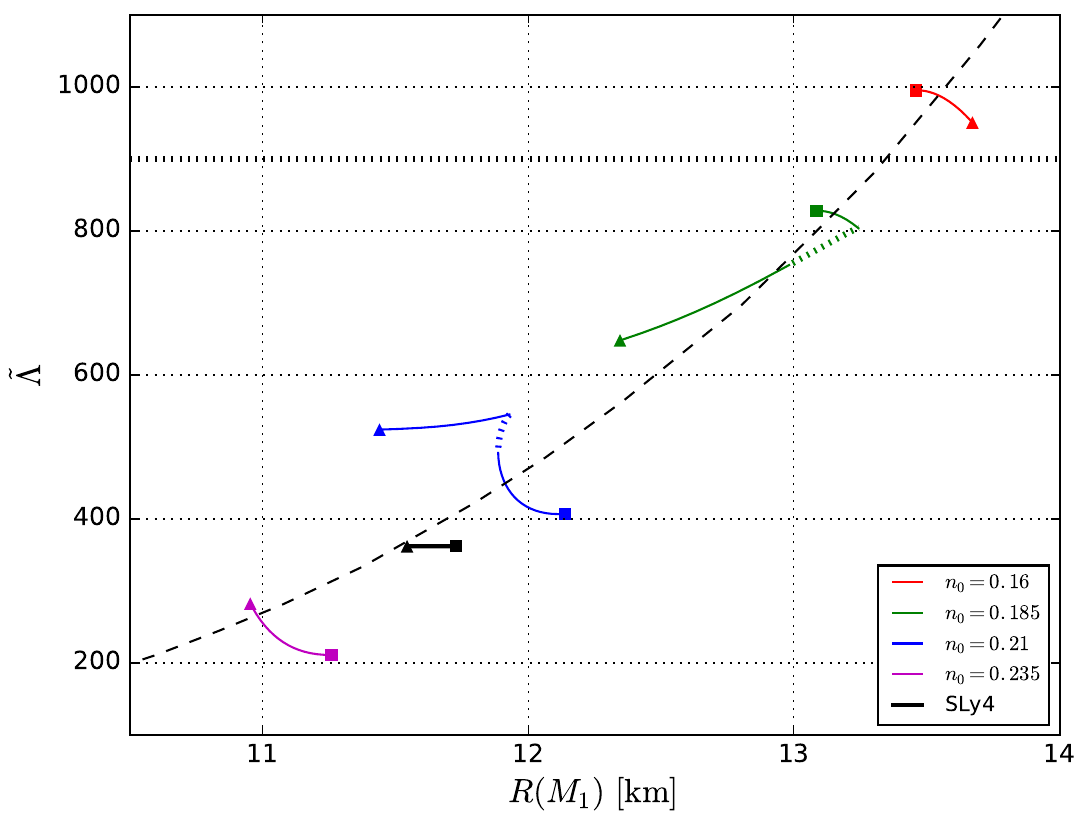}
\caption{$\tilde{\Lambda}$ as a function of the radius of the more massive
component, $R(M_1)$ for the low-spin estimation of parameters of the GW170817
event. Squares  denote the lowest estimated mass of the primary
component, $M_1=1.36\ M_\odot$, while triangles denote the highest mass,
$M_1=1.6\ M_\odot$. The dashed line corresponds to the fit by \citet{Raithel2018}.
The thick dotted line indicates the 90\% confidence limit of 900 for $\tilde{\Lambda}$ 
(corrected value from \citealt{Abbott2018b}, Table IV). Information about $\tilde{\Lambda}$ varying with the mass ratio $q=M_2/M_1$ for fixed chirp mass is encoded in the plot: $\tilde{\Lambda}$ curves for selected EOS follow the fixed chirp mass of GW170817. The square  denotes the $M_1=1.36 M_\odot$ (i.e. $M_1=M_2$, $q=1$), while the triangle denotes $M_1=1.60 M_\odot$ ($M_2=1.17 M_\odot$, $q=0.725$).} 
\label{fig:leffr1}
\end{figure}

Figure~\ref{fig:leffr1} shows the resulting $\tilde{\Lambda}$
(Eq.~\ref{eq:lambdatilde}) in a full range of the GW170817 event low-spin
component masses ($1.36-1.60\ M_\odot$ for $M_1$) as a function of the radius
$R(M_1)$ of the more massive component for the selected strong phase transition
EOS. These results are compared with the approximate formula for
$\tilde{\Lambda}$ proposed in \citet{Raithel2018}, based on a set of the
`standard' realistic EOS used in \citet{Abbott2017a}. The twin branch spans
a considerably wider range in $R$ for the allowed component masses of GW170817
than the standard EOS, e.g. the SLy4 EOS, for which the radii are
relatively constant. This feature results in a much broader range of allowed
$\tilde{\Lambda}$ values in comparison to the standard EOS: for comparison 
see Fig.~1 of \citet{Raithel2018}, where $\tilde{\Lambda}$ varies to a much
smaller extent than here. The behaviour of the red EOS (but also the violet EOS, in
which the phase transition occurs below the lower bound on the $M_2$ mass) is
qualitatively similar to the SLy4 EOS, but since the radii of the  components  change
more than in the case of the SLy4 EOS in the appropriate mass ranges, the
response of the star to the tidal deformation is larger, as the $\Lambda$
paramteter is a sensitive function of the radius. We expect a similar behaviour 
for any other EOS with a similar $M(R)$ sequence. The deviation from the fit is of
the order of 100 in $\tilde{\Lambda}$ and of several hundred  metres in
$R(M_1)$. The character of the curves depends on whether the phase transition
occurs within the range of $M_2$ or $M_1$ (blue and green lines). In the first
case, we note a discontinuity in $\tilde{\Lambda}$ related to the
phase-transition induced instability; in the second case, however, the $R(M_1)$
also behaves non-continuously because  a range of $M_1$ for
which two solutions with different values of  $R$ is possible.  

\subsection{Features of the quark twin branch for given polytropic outer layer}
\label{sect:outer}

In the following we present our study of the polytrope-quark matter EOS in order to connect 
the properties of the quark part of the $M(R)$ diagram to its less-dense, outer part 
i.e. the polytrope. The last $M(R)$
configuration of the polytropic EOS (corresponding to the transition point) 
is denoted by $C_{ph}$. For the
sake of presentation we  use a specific $C_{ph}$ configuration with 
$M_{ph}=2\, M_\odot$. It was selected because of its relation
to the current maximum observed mass, which we  assume to be equal
$M_{max}=2\,M_\odot$. 

\begin{table}
  \centering 
\begin{tabular}{| c | c | c | c | c |}
\hline
Model&$n_{0}$[fm$^{-3}$]&$\gamma$&$n_{1}$[fm$^{-3}$]&$\lambda$\\
\hline 
\hline
Dark Blue (DB) &0.192&5.45&0.325&1.9\\ 
\hline 
Blue (B) &0.1&3.7&0.37&1.9\\ 
\hline 
Cyan (C) &0.07&3.2&0.41&1.9\\ 
\hline 
Dark Red (DR) &0.3&5.0&0.455&1.6\\ 
\hline 
Red (R) &0.228&4.0&0.48&1.6\\ 
\hline 
Orange (O) &0.14&3.418&0.518&1.6\\ 
\hline 
Yellow (Y) &0.07&2.96&0.58&1.6\\ 
\hline
\end{tabular}
\vskip 1em 
\caption{Parameters of NSs indicated in Fig.~\ref{fig:alpha_gamma}. For all EOS speed of sound in quark matter ($\alpha$ parameter) was set to be equal 1.}
\label{tab_of_EOS}
\end{table}

\begin{figure}
   \centering
   \includegraphics[width=\columnwidth]{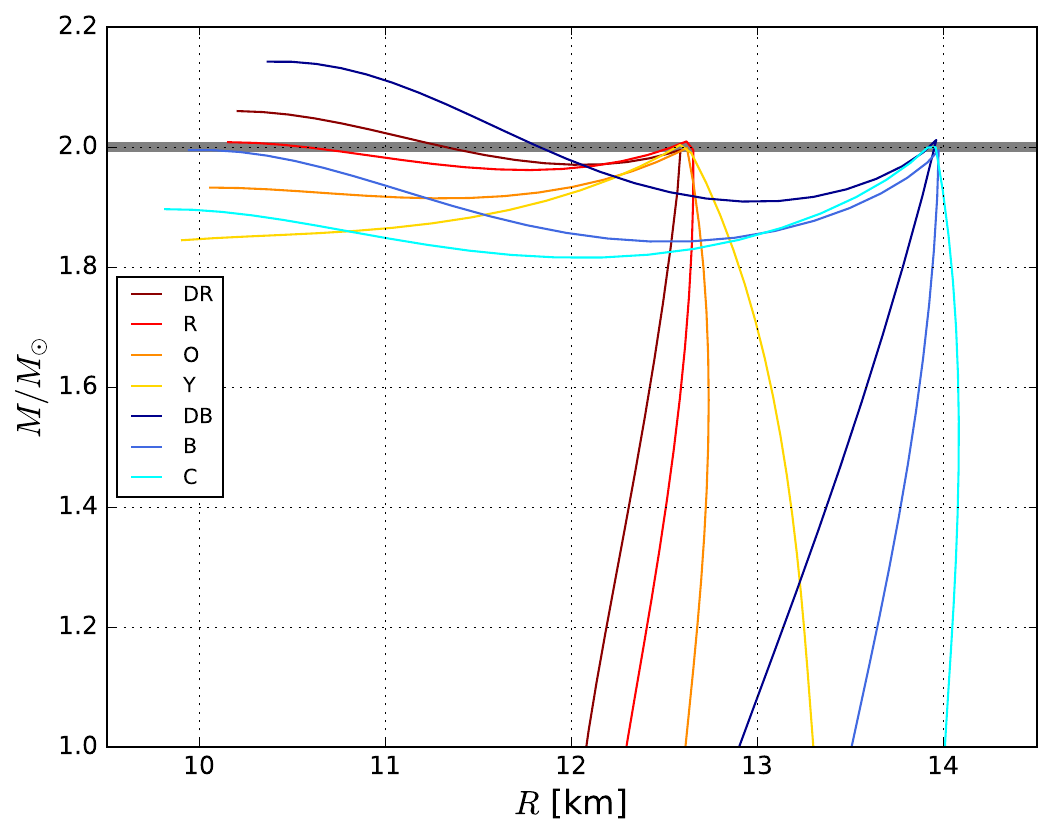}
      \caption{Example of the relation between the various EOS parameters and
properties of the twin branch. The state of matter at the phase transition
point (here quantified by the $dM/dR$ slope on the $M(R)$ plane) is related to the parameters of
the interior EOS. For the DB and DR EOS $M_{max}>M_{ph}$; for the R and B EOS $M_{max}=M_{ph}$; for the O and C EOS $M_{max}<M_{ph}$; for the Y EOS there are no stable twin solutions.}
         \label{fig:alpha_gamma}
   \end{figure}

\begin{figure}
   \centering
   \includegraphics[width=\columnwidth]{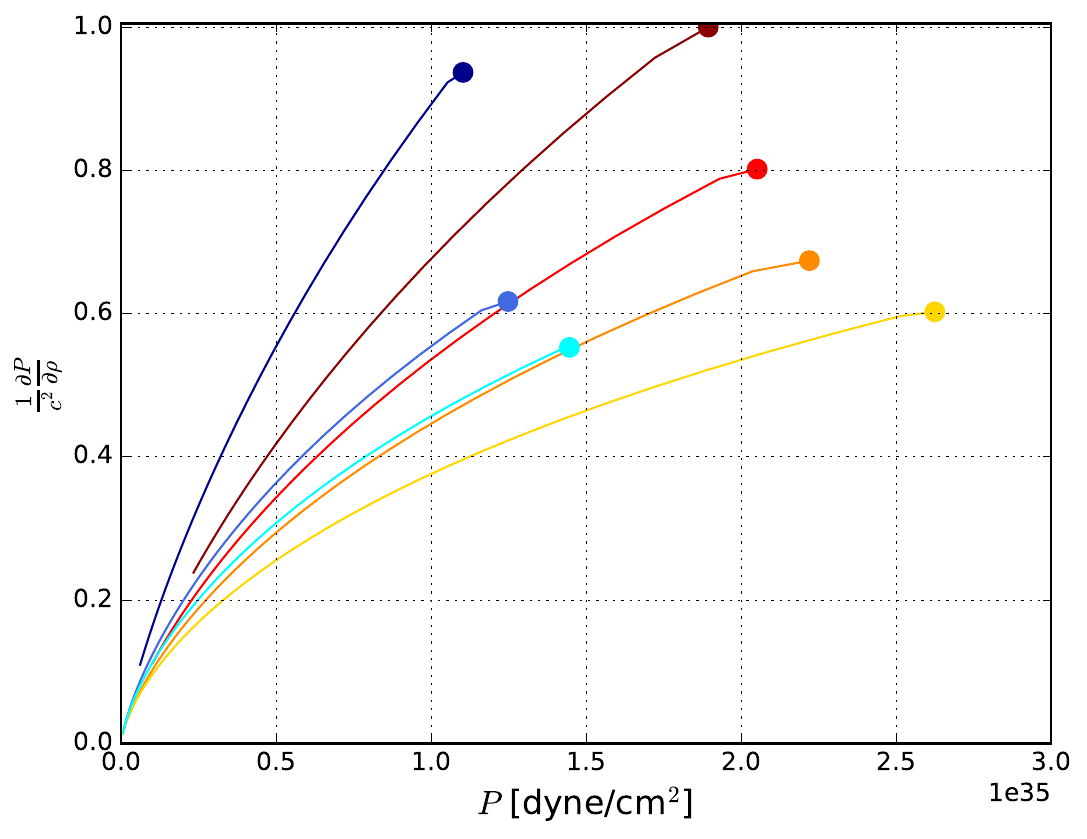}
      \caption{Relation between the square of the speed of sound $\alpha=\frac{\partial P}{\partial \rho}$ and pressure $P$ in the polytropic part of EOS (before the phase transition point) for the exemplary EOS introduced in Table~\ref{tab_of_EOS}. $M(R)$ relations produced for these EOS are shown in Fig.~\ref{fig:alpha_gamma}. }
         \label{fig:vs_P}
   \end{figure}

Figure~\ref{fig:alpha_gamma} shows a complimentary set of features. Here we
want to study the influence of the $C_{ph}$ configuration on the twin  branch.
We fix the $\alpha$ parameter to the maximum ($\alpha=1$). The stiffness of the polytropic
EOS  (related to the value of $\gamma$) is directly connected with the derivatives $dM/d\rho_c$ or $dM/dR$ at $C_{ph}$, and  has a direct impact on the presence of the twin branch.  Stiffer EOS at $C_{ph}$
(higher $\gamma$ resulting in higher value of $dR/dM$) favours the presence and size of the twin branch. Additionally, we investigate the point before the onset of phase transition. Figure~\ref{fig:vs_P} shows the $\alpha$ versus $P$ relation at the end of the polytope (related to the $n_1$ parameter from Table~\ref{tab_of_EOS} for each EOS). The colours correspond to the $M(R)$ relations in Fig.~\ref{fig:alpha_gamma}.
For illustrative purposes to investigate the conditions at which the stable twin
branch may be present, we select several mass-radius relations with $M_{max} >
M_{ph}$ (configurations denoted dark red, DR, and dark blue, DB), $M_{max} =
M_{ph}$ (red, R, and blue, B), $M_{max} < M_{ph}$ (orange, O, and cyan, C), and one sequence without a twin branch (yellow, Y).  The parameters of these
configurations are given in Table~\ref{tab_of_EOS}.


\begin{figure}
   \centering
   \includegraphics[width=\columnwidth]{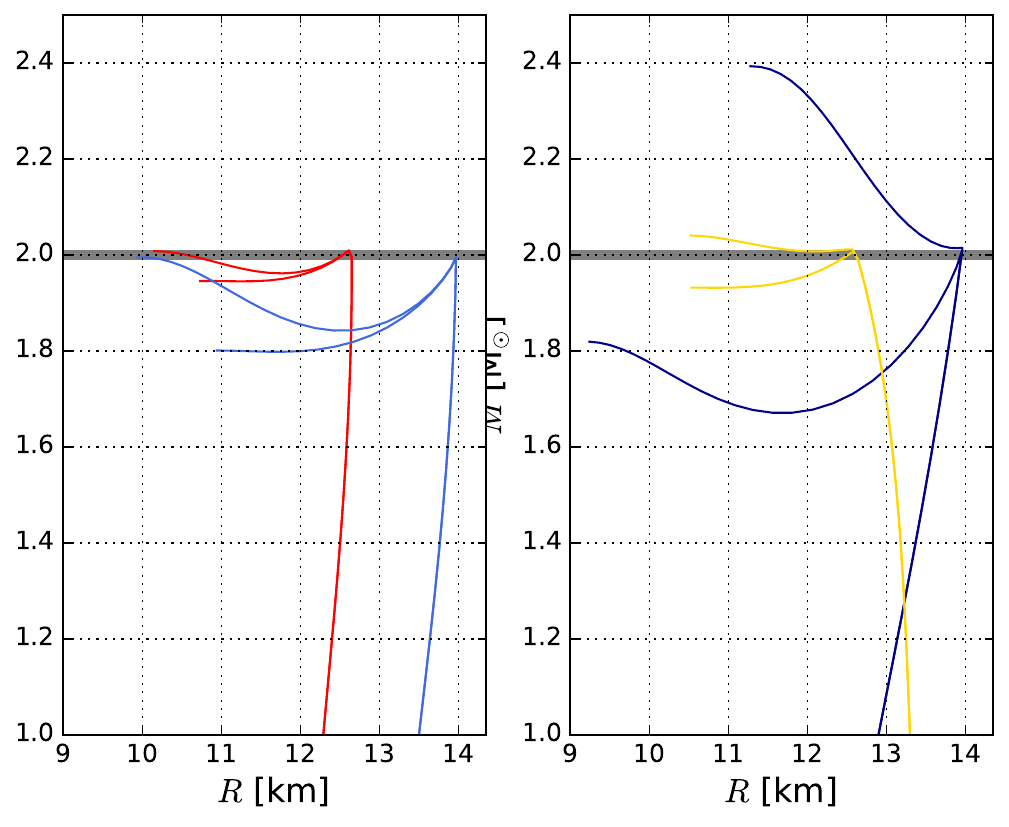}
      \caption{Presence  of a twin branch as a function the speed of sound in quark matter $\sqrt{\alpha}$  (blue and red curves, left panel) 
and the density jump $\lambda$ (dark blue and yellow,  right panel) 
for the exemplary EOS. Only configurations with the critical values of $\alpha$ and $\lambda$ for which a stable 
twin branch can be produced are shown. The high-mass twin branch segments have $\alpha$ in the range from 0.78 (bottom red curve) to 1.00 (top red curve) and 0.60 (bottom blue curve) to 1.00 (top blue curve). Values of $\lambda$ are in the range from 
1.58 (top dark blue curve) to 2.5 (bottom dark blue curve) and 1.34 (top yellow curve) 
to 1.48 (bottom yellow curve). The remaining EOS parameters ($\gamma$, $n_1$, $\lambda$, $n_0$ for the right panel and $\gamma$, $n_1$, $n_0$, $\alpha=1$ for the left panel)  correspond 
to the EOS in Table ~\ref{tab_of_EOS}.}
         \label{fig:alpha_stiff}
   \end{figure}

Figure~\ref{fig:alpha_stiff} demonstrates the behaviour of the $M(R)$ relations
for the selected values of the speed of sound parameter $\alpha$ (right panel)
and the density jump parameter $\lambda$ (left panel). Only critical values of
$\alpha$ and $\lambda$ parameters for which stable twin branches can be
formed are shown. As expected, the value of the sound velocity of a quark matter $\sqrt{\alpha}$ influences the stiffness of the
core and ability to stop gravitational collapse due to instability resulting from softening by the phase transition. On the other hand,  the stiffness $\lambda$ influences the size of the instability region between polytropic one-phase stars and hybrid stars with a quark core. For the EOS with smaller radii, at the point of
the phase transition, it is harder to produce a stable twin branch. This means that all twin solutions for the R and Y EOS occupy a similar area on the
mass-radius plane, while twin branches for B and DB span a much wider space. From
the observational point of view, according to  Figs.~\ref{fig:alpha_stiff}
and \ref{fig:lam_alfas}, observations of the NS with smaller radii  would put stronger constraints on the parameters of the possible phase transition to quark matter and the quark matter EOS, e.g. values of $\alpha$ and $\lambda$, as well as
the macroscopic properties of the NSs, e.g. the  minimum  radius $R_{min}(M_{max})$, 
corresponding to the maximum mass on the twin branch. 

The onset and parameters of the polytrope have a crucial influence on the radius
for configurations below $2\,M_\odot$ and on the behaviour of the twin branch. The
smaller the radius (the higher the value of $n_1$, the smaller the value of
$\gamma$), the higher the value of $\alpha$ needed to yield a stable twin branch (compare the lowest red and blue sequences in the left panel of
Fig.~\ref{fig:alpha_stiff}, where lower limit for the R EOS equals $\alpha=0.78$ for $\lambda=1.60$, and $\alpha=0.60$ for $\lambda=1.90$ for the B EOS).  A similar situation, but for the $\lambda$ parameter, is
presented in the right panel of the same figure: stable twin branches for the DB EOS can
be formed for values of $\lambda$ between 1.58 and 2.5, while for the Y EOS the allowed 
$\lambda$ range is much smaller, and lies between 1.34 and 1.48 (both for $\alpha=1$).

\begin{figure}
   \centering
   \includegraphics[width=\columnwidth]{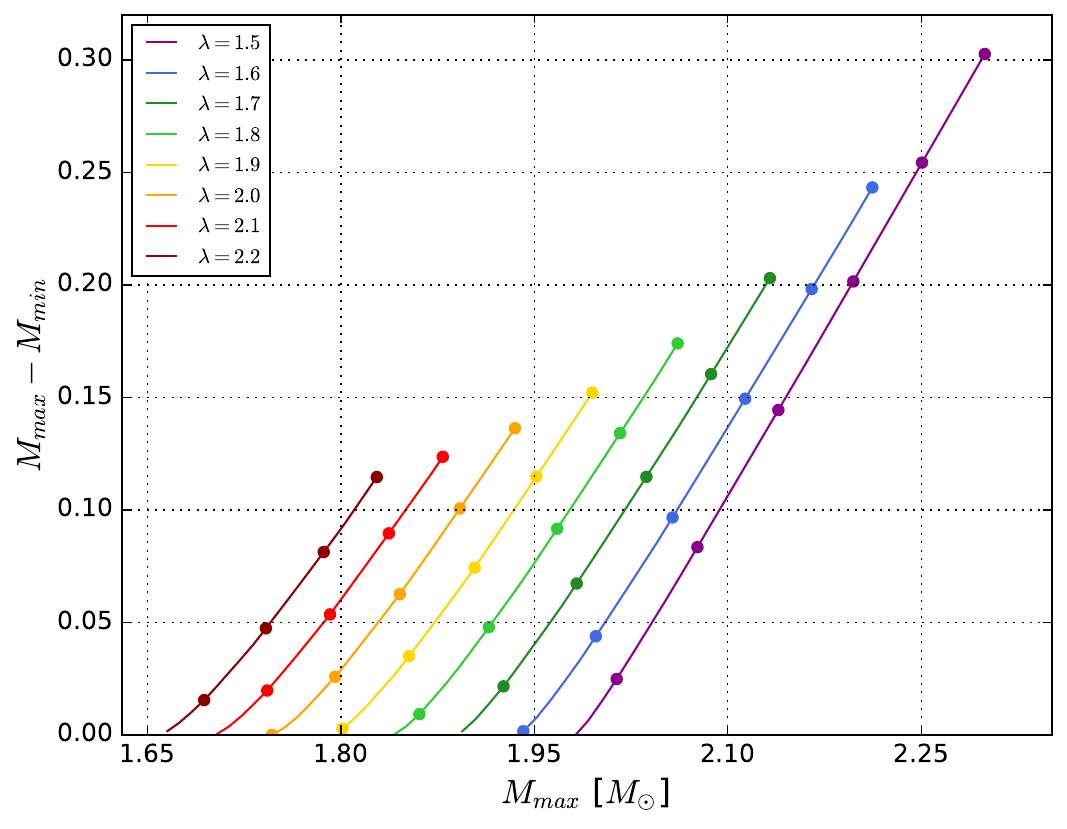}
   \caption{Exemplary B EOS case where the phase transition occurs at 
   $M_{ph}=2 M_\odot$. The relation between the values of maximum mass $M_{max}$  
   and the difference between $M_{max}$ and $M_{min}$. For a given $\lambda$ this dependence is approximately linear for a broad range of $\lambda$ parameter and $\alpha$ parameters. Along each curve the $\alpha$ parameter decreases from top to bottom if the figure, with a step of 0.1. The topmost dots correspond to $\alpha=1.0$.  }
         \label{fig:mdeltam}
   \end{figure}

\begin{figure}
   \centering
   \includegraphics[width=\columnwidth]{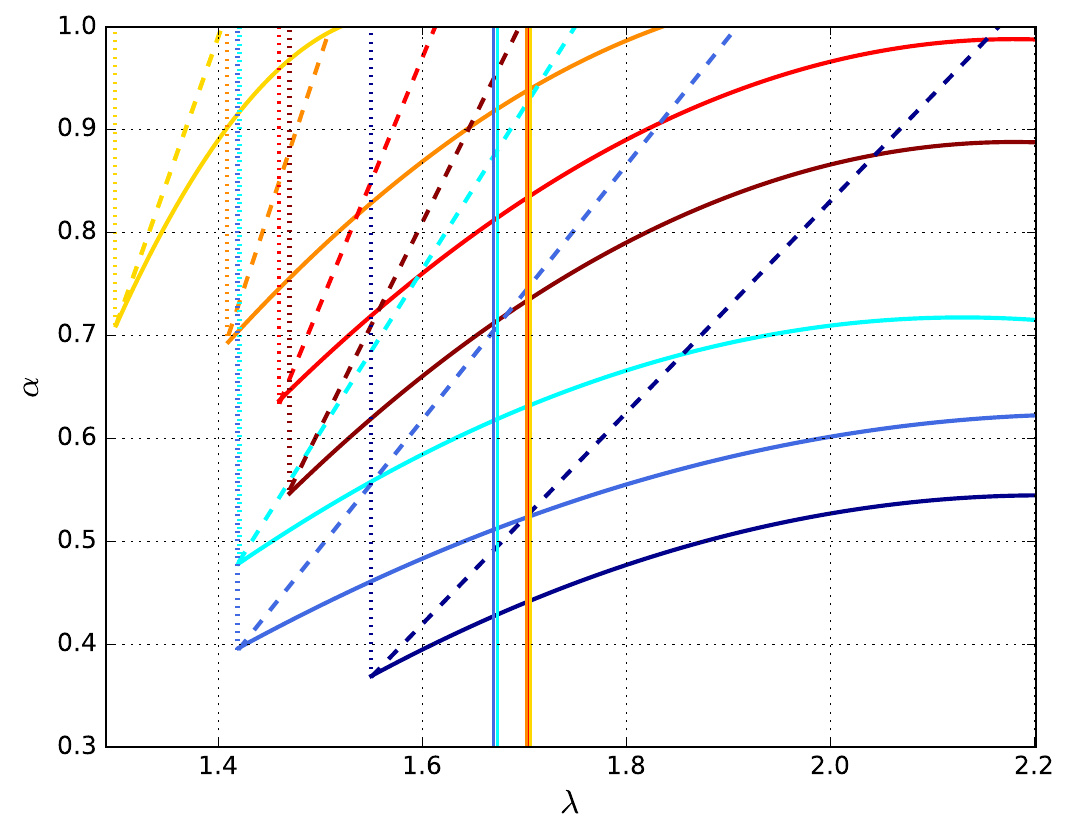}
   \caption{Ranges of $\lambda$ and $\alpha$  compatible with the 
the conditions of the presence of a stable second twin branch (above the solid curves) 
and $M_{max}>2\,M_\odot$ (above the long-dashed lines). The region between the dotted and solid vertical lines and the solid curve shows the region in which the solution denoted ``Both'' in \cite{Alford2013} are present. For the Y, O, R, and DR EOS the region is limited by the similar value of $\lambda\approx 1.7$, while for the C, B, and DB EOS $\lambda\approx 1.67$.}  
   \label{fig:lam_alfas}
\end{figure}

Figure \ref{fig:mdeltam} illustrates the conditions for a stable twin branch.
For this, local mass minimum and maximum, $M_{min}$ and $M_{max}$, is required.
A marginal case corresponds to $M_{min}=M_{max}$. These configurations
correspond to the zeros of the approximately linear relation between $M_{max}$
and the difference $M_{max}-M_{min}$ in Fig.~\ref{fig:mdeltam} for the values
of $n_0$, $n_1$, and $\gamma$ as for the B EOS. We note that for smaller values of $\lambda$: i) a smaller value of $\alpha$ is required to produce a stable twin branch, ii) higher values of $M_{max}$ are reached, iii) $(M_{max} - M_{min})$ obtains higher values.

For a given $C_{ph}$ configuration we can plot a $\lambda(\alpha)$ relation, defining the onset of a twin branch, from
the crossing point of the curve plotted in Fig. \ref{fig:mdeltam} with
a horizontal line (y=0). For a given $\lambda$ this condition defines 
the critical value of $\alpha$. For the presence of a new, stable twin branch,
matter in a dense phase should be sufficiently stiff: $\alpha>\alpha_{crit}$.

\begin{figure}
\centering
\includegraphics[width=\columnwidth]{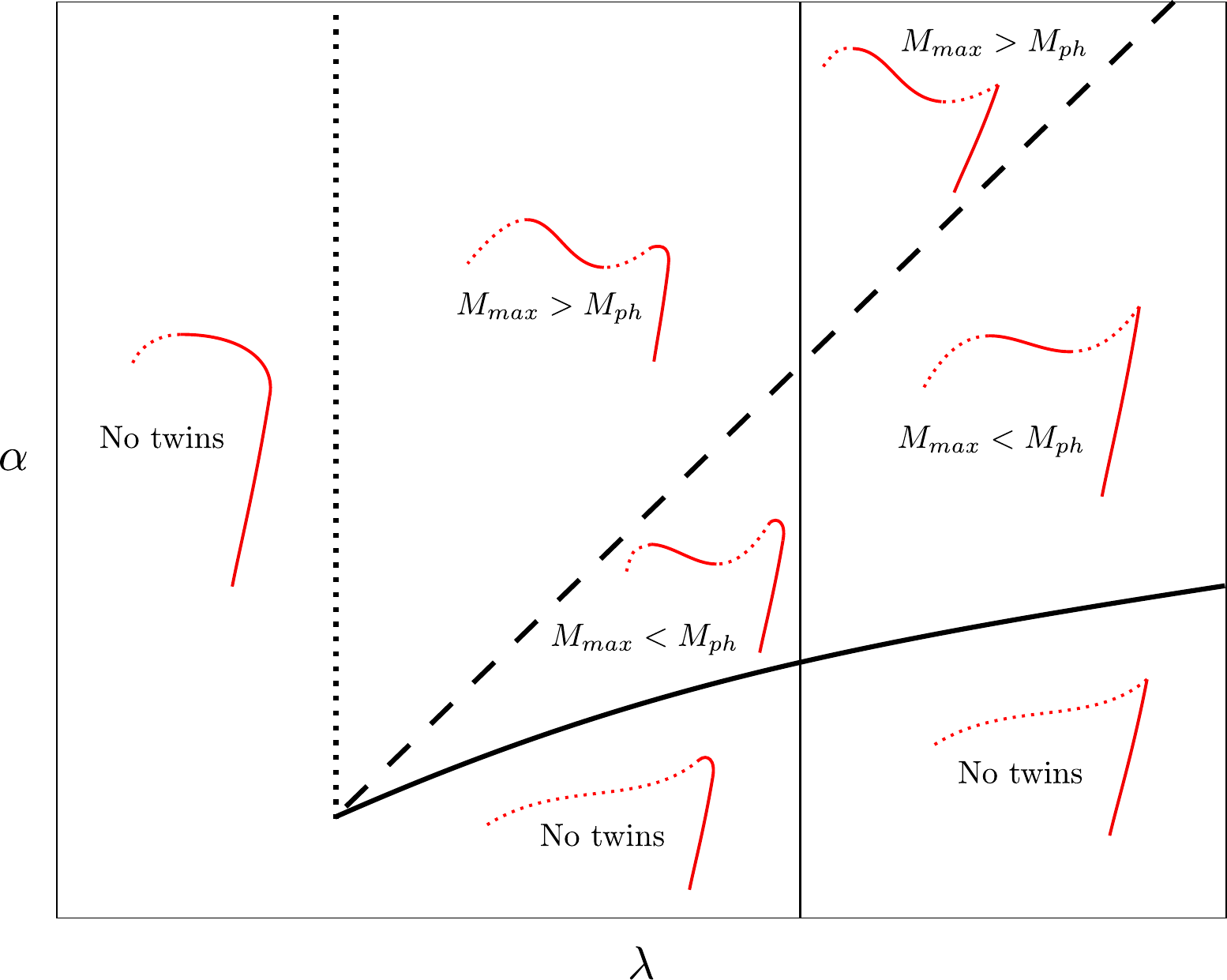}
\caption{Schematic explanation of Fig.~\ref{fig:lam_alfas}. 
Above a critical $\lambda$ (above the solid line) twin solutions 
are possible. With increasing $\alpha$, the mass on the twin branch may be higher 
than the mass at the phase transition point $M_{ph}$ (regions above the long-dashed 
line). The region between the vertical dotted and solid lines corresponds to the ``Both'' class, 
described in \citet{Alford2013}, which is characterised by a delayed onset of 
instability with respect to configurations with larger $\lambda$.} 
  \label{fig:alpha_delta_types}
\end{figure}

Another astrophysically important condition corresponds to the presence of stellar configurations
on the twin branch with baryon masses higher than the maximum mass of one-phase star $M_{ph}$.
In this case there is a natural evolutionary track which allows for the
creation of twin stars. The condition $M_{B,max}=M_{B,ph}$ defines for each $\lambda$ the  limiting
value of $\alpha$ - $\alpha_{Mmax}$.
This condition is stronger than just the presence of stable twin branch  and
$\alpha_{Mmax}>\alpha_{crit}$.
Strictly speaking this condition corresponds to the comparison of {\it baryon} masses. However, the 
relative difference in the gravitational mass of normal and twin sequence for the same $M_B$ is of the order of ${\sim}10^{-3}$, 
and within this accuracy we can consider the condition $M_{max}=M_{ph}=2M_\odot$, which is equivalent to the crossing
points of curves presented in Fig. \ref{fig:mdeltam} with vertical line $M_{max}=2\,M_{\odot}$. Conditions for the $\alpha$ and $\lambda$ parameters for which a stable twin branch is possible are shown in Fig. \ref{fig:lam_alfas}. 
A schematic explanation of this plot is presented in Fig. \ref{fig:alpha_delta_types}. 
Stable twin solutions are not possible to the left of the dotted vertical lines, nor below the solid curves. Between the solid curves and the dashed lines, a stable hybrid branch can be formed with $M_{max} < 2M_{\odot}$. Dashed lines correspond to the situation in which $M_{ph}=M_{max}=2M_{\odot}$, 
while the regions between the dashed and dotted curves correspond to $M_{max}>M_{ph}$. 

We also indicate a region of the  ``Both'' class proposed by \citet{Alford2013}: it lies between the vertical dotted and vertical solid lines. These configurations are not immediately destabilised after the phase transition;  both maxima in the mass-radius relation occur for a finite size quark-phase core. For the Y EOS, all twins are of the ``Both'' class. Limiting $\lambda$ for the DB, B, and C EOS at which ``Both'' configurations can still be formed ${\approx}1.67$, while for the Y, O, R, and DR EOS it equals ${\approx}1.70$. A clear separation between these two types of EOS is, in this context, the effect of the separation of the radii at the phase-transition point.

Parameter $n_{1}$ decides where the transition point $C_{ph}$ occurs. Here we
investigate the properties of the twin branch with respect to the density at which
the polytrope ends.

\begin{figure}
\centering
\includegraphics[width=\columnwidth]{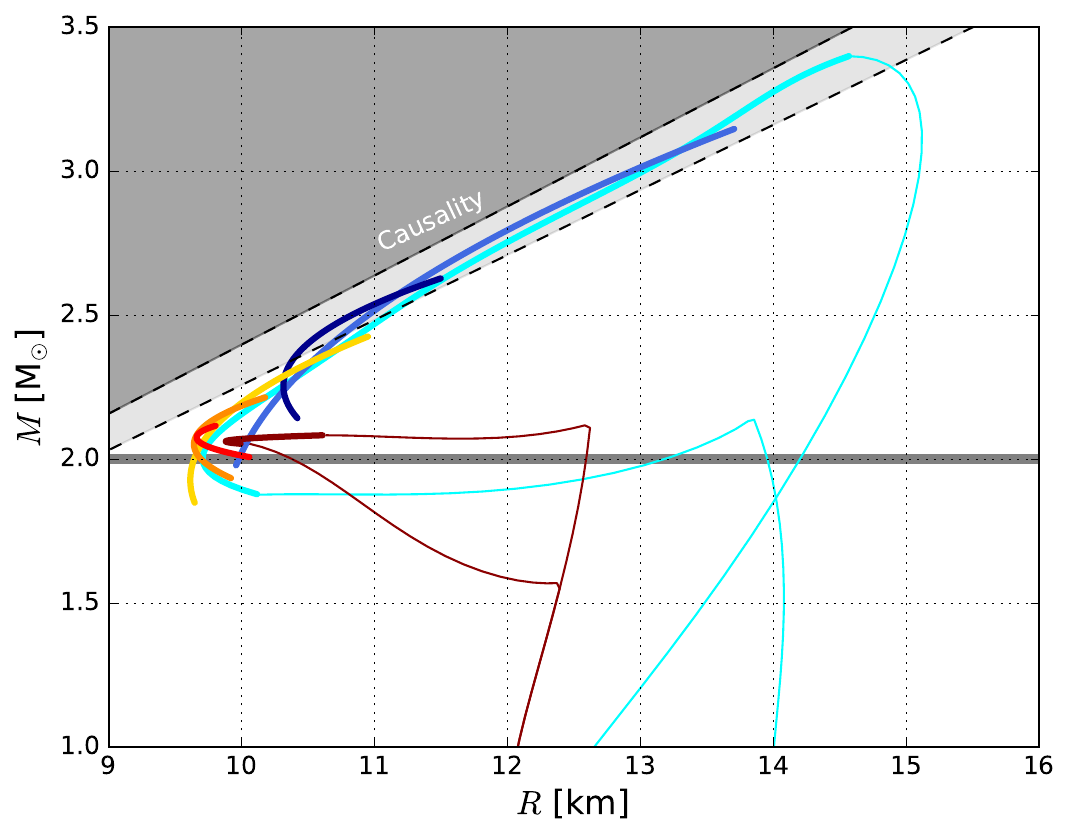}
\caption{Mass-radius relations for the marginal values of $n_{1}$, for which stable twin branches can be formed, in the case of the DR and C EOS (colours correspond to the same parameters, except $n_1$ values, as shown in Fig.~\ref{fig:alpha_gamma} and in Table ~\ref{tab_of_EOS}). Thick lines correspond to the $M_{max}$ (or $R_{min}$ on the twins branches) for all ranges of $n_1$. The causal limit is shown as a dark shaded area, when the light shaded space corresponds to the photon orbit in the Schwarzschild case. Some stable configurations can be formed within the photon orbit. For the C EOS, when $M_{max}$ is high, the quark phase occurs for very low masses.} 
  \label{fig:nb_rmin}
\end{figure}

\begin{figure}
\centering
\includegraphics[width=\columnwidth]{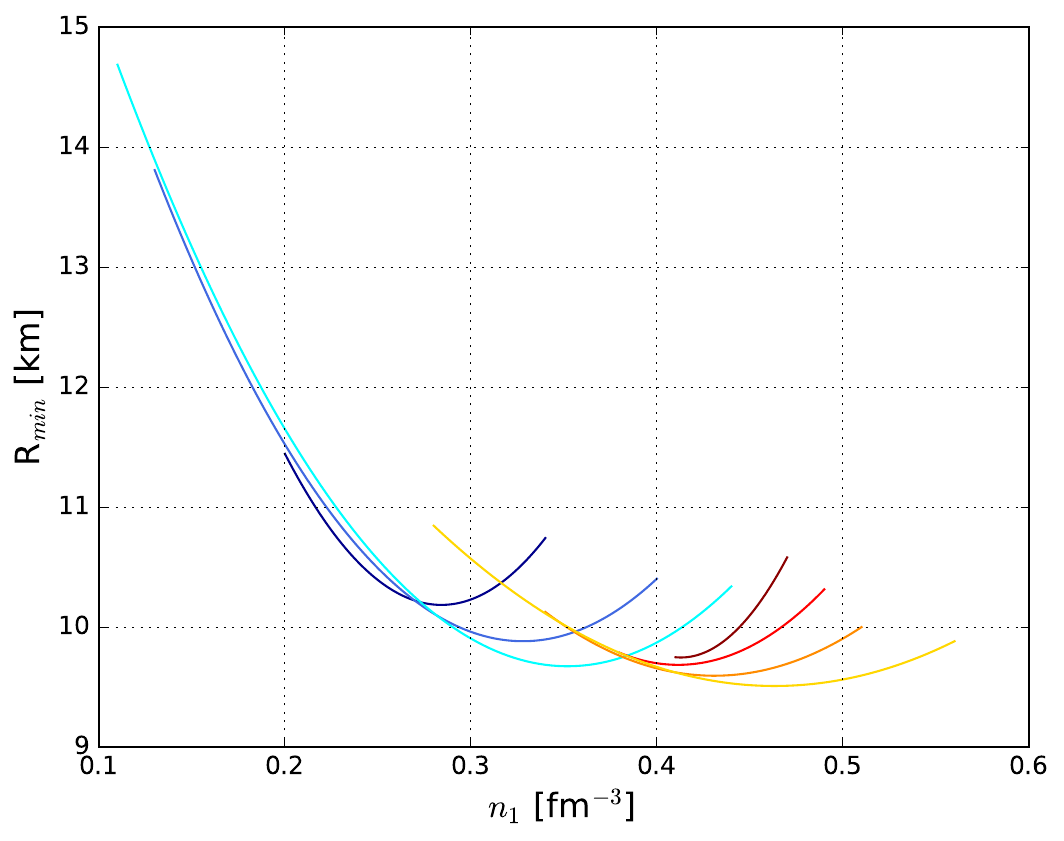}
\caption{Minimum radii $R_{min}$ (thick lines in Fig.~\ref{fig:nb_rmin}) as a function of baryon density where the phase transition occurs $n_1$.  }
  \label{fig:rnb}
\end{figure}

For the set of EOS from Table~\ref{tab_of_EOS}, we changed
parameters $n_{1}$ and chose only the configurations with stable twin branches, as
shown in Fig.~\ref{fig:nb_rmin}. The configurations with larger radii at the polytropic 
branch (the C, B, DB EOS) span a wider range of $n_{1}$ 
for which our assumptions are fulfilled (see also Fig.~\ref{fig:rnb}). 
This kind of EOS support situation, where the phase transition point occurs
at low $n_1$ and where very massive, stable NSs (with $M_{max}$ even higher than $3
M_{\odot}$) can exist on the quark branch. We note that all possible
$R_{min}(M_{max})$ for the DR EOS are in the range $\approx 9.8-10.5$ km, while for the C
EOS this range is much wider: $\approx 9.7-14.8$ km (shown in
Fig.~\ref{fig:nb_rmin} as thick lines and in Fig.~\ref{fig:rnb} as a
function of $n_1$). We compare these results with the causal limit (see e.g.
Appendix D in \citealt{HPY2007book}):
\begin{equation}
\frac{2GM}{Rc^2}\approx 0.71.  
\label{eq:causal_M}
\end{equation}
The maximum mass $M_{max}$ is then related to the density $\rho_{*}$ as
\begin{equation}
M_{max} = 2.116 \cdot (\rho_{*,15})^{-1/2}~M_{\odot}, 
\label{eq:causal_Mmax}
\end{equation}
where $\rho_{*,15} = \rho_{*}/10^{15}$ g/cm$^{3}$ (see Eq. \ref{eq:bag}). 
The $M(R)$ relations for all the EOS are limited by Eq.~\ref{eq:causal_M} and asymptotically converge to the causal limit 
with decreasing $n_1$.

Minimum radii corresponding to $M_{max}$ are indicated  by thick lines in Fig.~\ref{fig:nb_rmin}, with $R_{min}$ between 9.5 and 10.5 km, depending on the EOS. For low values of $n_1$ the radii decrease approximately along the causal limit, reach a value of $R_{min}$, and become bigger again for high values of $n_1$. 

\begin{figure}
\centering
\includegraphics[width=\columnwidth]{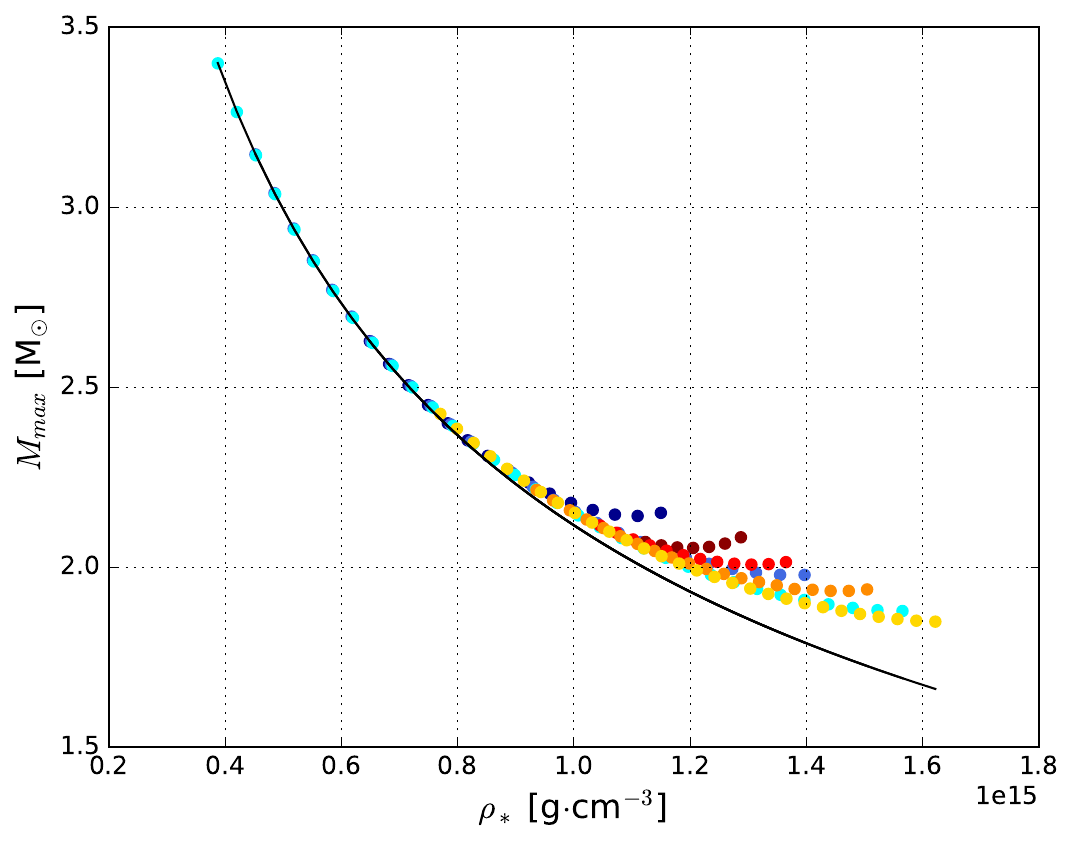}
\caption{Maximum mass $M_{max}$ on the twin branch vs. $\rho_{*}$ (see Eq.~\ref{eq:bag}). The causal limit (Eq.~\ref{eq:causal_Mmax}) is shown by a black line.} 
  \label{fig:mph_rho}
\end{figure}

We also investigate our results in the context of Eq.~\ref{eq:causal_Mmax}, as  shown in Fig.~\ref{fig:mph_rho}. For the configurations with dominant contribution of the quark phase (like the C EOS), the results converge to the causal limit. If the phase transition starts at higher densities (corresponding to $\rho_{*}\approx 0.9\times 10^{15}$ g/cm$^3$), calculated curves diverge from the causal limit. The divergence is especially strong for the O, R, and DR EOS, for which twin branches can be formed for relatively small radii at the transition point. According to the recent analysis of the GW170817 event in \cite{Abbott2018a}, these EOS are more favourable than C, B, and DB, due to the smaller radii for configurations without phase transition in the centre  and with masses relevant to GW170817 ($M\sim 1.2 - 1.6 M_{\odot}$).

On the twin branch the minimum radius of the hybrid star  depends on the density at phase transition $n_1$ (see Fig.~\ref{fig:rnb}). For the adopted EOS models the maximum $n_1$ for which a stable, twin branch exists is about $0.56$ [fm$^{-3}$].

\section{Conclusions}
\label{sect:concl}

In our work we consider stable NSs configurations, described by the EOS constructed  using the SLy4 prescription of the crust, joined with the relativistic polytrope, which ends in the transition point where the quark matter phase occurs. This internal structure may reflect as a twin branch in the mass-radius relation, which means that objects with  different $R$ might have the same $M$.

Recent detection and improved analysis of the gravitational wave event, originated as a NS-NS merger and named GW170817, delivered information about tidal deformabilities of the compact objects. By using the multiple hybrid EOS, we show that the behaviour of the $\Lambda(M,R)$ is not straightforward, as regions with different $\Lambda$ values occupy the same space in the $M(R)$ plane. We point out that representing $\Lambda(M_{1.4})$ on the $M(R)$ plane, as  was recently used in the literature, is not optimal for examining the $M(R)$ dependency of the tidal deformability on the EOS. We also show that the non-monotonic behaviour of $\Lambda_1 - \Lambda_2$ relation depends on the size of the density jump $\lambda$, and directly reflects the non-monotonic behaviour of $M(\Lambda)$ function. An even more complicated situation occurs when the dependency on the SLy4 crust-polytrope connection point is included.

We find that objects on the twin branch are more compact (in comparison to the polytropic branch) and give smaller $\Lambda$ values. Such solutions are more favoured from the point of view of the GW170817 analysis. In the case of a high-mass twin branch (where the transition point occurs around $2~M_{\odot}$), stars in the range of GW170817 masses have larger $R$ and larger $\Lambda$ than lower-mass twins. Because the NS with the same compactness can have different tidal deformability (due to the different EOS), we show that the deviation from the $\Lambda - GM/Rc^2$ fit proposed in the literature, for the broad range of the hybrid EOS, is around $5\%$ for $\Lambda < 100$ and increases to above $10\%$ for the higher values of $\Lambda$. We find a similar inaccuracy in the fit on the $\tilde{\Lambda}$ versus $R(M_1)$ plane, where discontinuous relations for the hybrid EOS span a wider range in $R$ than the standard ones (e.g. the SLy4 EOS).

Finally, we focus on the influence of the transition point on the twin branch. Approximately linear relations for $M_{max} - M_{ph}$ versus $M_{max}$ (for a fixed $\lambda$ and changeable $\alpha$) are used to establish conditions for a density jump and speed of sound, that allow for the creation of different types of twin configurations (i.e. ``Both'' NSs). For the exemplary EOS, most of the twins with $M_{max}>M_{ph}$ can be formed for at least $\lambda = 1.33-1.54$ and corresponding values of $\alpha = 0.7 - 0.37$, while ``Both'' stars occupy the $\lambda \lesssim 1.7$ range. Generally, the smaller the $R$ on the polytropic branch, the more fine-tuned the conditions need to be  to produce a stable twin branch. Our whole set of solutions asymptotically converges to the limit of the causal EOS with  decreasing values of $n_1$, and it can be approximated by the causal limit when the central density at the beginning of the twin branch $\rho_{*} \lesssim 0.9\times 10^{15}$ g/cm$^3$ (when contribution to the EOS of the quark phase is dominant). We also show that the minimum radius on the twin branch has a size between 9.5 and 10.5 km and depends on the $n_1$ value. We note that analyses were performed for exemplary EOS, and at most two varying parameters at once from the parameter set ($n_0$, $n_1$, $\gamma$, $\alpha$, $\lambda$), when other parameters were constant. This strategy simplified investigation and allowed us to determine the constraints mentioned above. Nevertheless, a full analysis is needed and is dedicated to  future studies. 

\begin{acknowledgements}
We acknowledge support from the Polish National Science Centre via grant nos.
2014/13/B/ST9/02621, 2014/14/M/ST9/00707, and 2016/22/E/ST9/00037, and from the
European Union’s Horizon 2020 research and innovation programme under grant
agreement no. 653477. WT acknowledges the support of the Nicolaus Copernicus
Astronomical Center 2015 summer studies programme.
\end{acknowledgements}


\end{document}